\documentclass[pre,preprint, showpacs]{revtex4}
\usepackage{amsmath}
\usepackage{dcolumn}
\usepackage[xdvi]{graphicx}%
\makeatletter

\def\btt#1{\texttt{\@backslashchar#1}}%
\DeclareRobustCommand\bblash{\btt{\@backslashchar}}%
\makeatother

\newcommand{\bv}[1]{{\boldsymbol #1}}

\begin{document}

\title{Spatial correlations in sheared isothermal liquids : From elastic particles to granular particles  }

\author{ Michio Otsuki$^1$$^{,2}$\footnote{Present address.} and Hisao Hayakawa
$^1$}
\affiliation{$^1$ Yukawa Institute for Theoretical Physics, Kyoto University,  Kitashirakawaoiwake-cho, Sakyo-ku, Kyoto 606-8502, JAPAN.\\
$^2$ Department of Physics and Mathematics, Aoyama Gakuin University,
5-10-1 Fuchinobe, Sagamihara, Kanagawa, 229-8558, JAPAN }

\begin{abstract}
Spatial correlations in sheared isothermal liquids for both elastic and granular cases are theoretically investigated.
Using the generalized fluctuating hydrodynamics, correlation functions for both the microscopic scale and the macroscopic scale 
  are obtained. We find the existence of  long-range correlations obeying power laws. 
The validity of our theoretical predictions has been verified from the molecular dynamics simulation.
\end{abstract}
\pacs{83.50.Ax, 45.70.-n, 61.20.-p, 45.50.-j}

\maketitle

\section{Introduction}
Liquids consist of assemblies of many particles. 
There is long history to study molecular liquids which consist of non-dissipative simple molecules \cite{Hansen,Zoppi}.
In these days, there has been rapid growing interest in granular liquids which are made of 
granular assemblies, i.e., dissipative macroscopic 
particles \cite{aronson,pouliquen,silbert,lutsko01,alam03,GDRMiDi,lutsko04,namiko05,kumaran,orpe07,hisao07a,namiko07,hisao07b,Otsuki,hayakawa08,kudrolli}.
Although a molecular liquid is fluctuated around an equilibrium state, 
the absence of the equilibrium state of granular assemblies makes granular materials
unlike usual materials \cite{jaeger}.
Nevertheless, we have recognized that statistical properties and hydrodynamic behaviors of moderate dense 
and nearly elastic granular gases are considerably understood 
from the analysis of the kinetic theory \cite{lutsko04,hisao07a,Brilliantov,JR85a,JR85b,Garzo,Lutsko05}.

Boltzmann-Enskog theory is often used in describing moderate dense gases including
 granular assemblies, where the assumption of molecular chaos is used \cite{lutsko04,Brilliantov,JR85a,JR85b,Garzo,Lutsko05}.
In spite of the success of qualitative description of hydrodynamic  behaviors
 based on Boltzmann-Enskog theory,  
significant roles of correlations have been recognized in these days. 
Indeed, it is well known that correlated collisions cause significant
differences in constitutive equations even in equilibrium gases. 
In particular, recently a number of papers on long-time tails in current correlation functions for granular liquids 
have been published, which are directly related to 
the transport coefficients \cite{kumaran,orpe07,hisao07b,Otsuki,kudrolli,hayakawa08b,otsuki08b,kumaran08}.
On the other hand, we know the existence of long-range correlations 
 in sheared elastic fluids \cite{Lutsko85,Lutsko02,Wada}
 and heat conduction systems \cite{sengers1,sengers2,sengers3,sengers4,dorfman,sengers5,sengers6}.
 The existence of a similar long-range
correlation has been observed even in a simulation of sheared granular fluids \cite{otsuki08a}.
In spite of the indication of the existence of the long-range correlations in sheared fluids,
 we do not have any consistent theory in describing a structure factor for 
 both the particle scale and the hydrodynamic scale.
Indeed, a theoretical prediction of spatial correlations 
in the molecular scale for relatively dense liquids under a shear is not 
consistent with the long-range correlations \cite{lutsko02b}.  
Thus, we still do not understand the details of spatial correlations even in the case of molecular liquids. 

We often use the mode-coupling theory (MCT) in describing dense liquids
\cite{gotze,das,reichman}. 
Even when we are interested in sheared dense granular liquids, the framework
of MCT can be used \cite{hayakawa08}. However, MCT is not a closed theory,
because it needs to determine the structure factor by another method.
Thus, to obtain the structure factor or the pair-correlation function is
an important issue for the description of dense granular liquids.

The purpose of this paper is, thus, to clarify the spatial correlation functions for  sheared isothermal liquids.
Based on the generalized fluctuating hydrodynamics \cite{Kirkpatrick85,Kirkpatrick86}, 
we will demonstrate that the long-range correlation obeying a power law is consistent
with the short-range structure obtained by a liquid theory.
In the next section, we will summarize the outline  of the generalized fluctuating hydrodynamics.
In Section III,  we will show properties of a set of linearized equations around a uniform shear flow (USF) based on the generalized fluctuating hydrodynamics.
In Section IV, we will calculate the spatial correlation functions and their asymptotic forms in the long-range limit.
In Section V, we will compare our results with those of the molecular dynamics simulation. 
In Section VI, we will discuss and conclude our results.
In Appendix A, we summarize the form of the pair-correlation function $g_0(r,e)$ for homogeneous unsheared states
obtained by Lutsko \cite{lutsko01}.
 In Appendix B, we briefly summarize the transformation between the  Cartesian coordinate and the oblique coordinate.
In Appendix C, 
we present the  explicit form of $\tilde{C}_{nn}(\bv{k})$
which is the Fourier component of the density correlation.
In Appendix D, 
we show the explicit form of 
 $\tilde{C}_{pp}(\bv{k})$
which is the Fourier component of the momentum correlation.
In Appendix E, we evaluate the asymptotic form of a function $\bar{\Delta}_j(r)$ which is needed for the calculation
of correlation functions. 


\section{Generalized Fluctuating Hydrodynamics}



We consider three-dimensional systems consisting of $N$ 
identical smooth and hard spherical particles confined in the volume $V=L^3$  
under a shear flow with the shear rate $\dot\gamma$, where each particle has the mass
$m$ and 
the diameter $\sigma$. 
The particles collide instantaneously with each other
by a restitution constant $e$ which is less than unity for granular particles, and  is equal to unity 
for molecular liquids.
Let us assume that the restitution coefficient $e$  is a constant.

The spatial correlations
in sheared fluids are investigated by using the fluctuating hydrodynamics \cite{Lutsko85,Lutsko02,Wada,otsuki08a}
which can be used for the description of hydrodynamic behaviors.
However, it is possible to generalize  the fluctuating hydrodynamics which can cover
the scale
around the particle diameter. 
The generalized hydrodynamic equations \cite{Kirkpatrick85,Kirkpatrick86,Das90,Marchetti} 
for isothermal liquids characterized by a uniform temperature $T$ are given by
\begin{eqnarray}\label{mass}
\partial_t n + \bv{\nabla} \cdot (n \bv{u}) & = & 0, \label{n:eq} \\
\partial_t u_\alpha + u_\alpha \nabla_\beta u_\beta + \frac{1}{m} \nabla_{\alpha}
\mu
+ \frac{1}{mn} \nabla_\beta (\Sigma_{\alpha \beta}^D + \Sigma_{\alpha \beta}^R)
& = & 0,  \label{u:eq} 
\end{eqnarray}
where  $n$ and $u_{\alpha}$ 
are  the number density and $\alpha$-component of the velocity field, respectively.
Here, we have introduced the generalized chemical potential or
the effective pressure
\begin{eqnarray}
\mu= T \left [ \ln n - \int d\bv{r}' C(\bv{r} - \bv{r}',e,\dot\gamma)
\delta n(\bv{r}',t) + \cdots
\right ],
\end{eqnarray}
where $C(\bv{r}-\bv{r}',e,\dot\gamma)$ is the two-particle direct correlation function, which satisfies
$nC(k,e,\dot\gamma) \equiv 1 - S(k,e,\dot\gamma)^{-1}$ with the structure factor $S(k,e,\dot\gamma)$.
$\Sigma_{\alpha \beta}^D(\bv{r},t)$ is the viscous stress tensor given by
\begin{eqnarray}\label{noise}
\Sigma_{\alpha \beta}^D(\bv{r},t) &=& -\int d \bv{r}' [ 
\eta(\bv{r}-\bv{r}',e)
\left \{ 
\dot{\epsilon}_{\alpha \beta}(\bv{r}',t)+\dot{\epsilon}_{\beta \alpha}(\bv{r}',t)-
\frac{2}{3}\dot{\epsilon}_{\gamma\gamma}(\bv{r}',t)\delta_{\alpha \beta}
\right \} \nonumber\\
& &
-\zeta(\bv{r}-\bv{r}',e)\dot{\epsilon}_{\gamma\gamma}(\bv{r}',t)\delta_{\alpha \beta} ],
\end{eqnarray}
where $\dot{\epsilon}_{\alpha \beta}(\bv{r},t)\equiv
\{ \nabla_\alpha u_\beta(\bv{r},t) + \nabla_\beta u_\alpha(\bv{r},t) \} /2$.
$\Sigma_{\alpha \beta}^R(\bv{r},t)$ is the random part of the stress tensor satisfying 
$\left < \Sigma_{\alpha\beta}^R(\bv{r},t) \right >=0$, and
\begin{equation}\label{noise-corr}
\left < \Sigma_{\alpha\beta}^R(\bv{r},t)  \Sigma_{\gamma\delta}^R(\bv{r}',t')\right >=
2T \delta(t-t') \{ \eta(\bv{r}-\bv{r}',e) \Delta_{\alpha\beta\gamma\delta} + \zeta(\bv{r}-\bv{r}',e) 
\delta_{\alpha\beta} \delta_{\gamma\delta} \}
\end{equation}
 with $\Delta_{\alpha\beta\delta\gamma}
 \equiv \delta_{\alpha\gamma}\delta_{\beta\delta} + 
\delta_{\alpha\delta} \delta_{\beta\gamma} - 2\delta_{\alpha\beta}\delta_{\gamma\delta}/3$.
Here, we have used Einstein's sum rule on the Greek subscript. 
The generalized shear viscosity $\eta(\bv{r},e)$ and the generalized bulk viscosity $\zeta(\bv{r},e)$ are
represented by $\nu^*_1(k,e)$, $\nu^*_2(k,e)$ and Enskog's mean free time 
\begin{equation}\label{enskog_time}
t_E\equiv \frac{1}{4\pi n_0\sigma^2 g_0(\sigma,e)}
\left(
\frac{m\pi}{T} 
\right)^{1/2}  
\end{equation}
  with
the radial distribution function $g_0(\sigma,e)$
at contact
 as  $\nu^*_1(k,e) = (mn_0 \sigma^2 t_E^{-1})^{-1}
(\zeta(k,e) + 4\eta(k,e)/3)$, and $\nu^*_2(k,e)= (mn_0 \sigma^2 t_E^{-1})^{-1}
\eta(k,e)$, where $n_0$, $\eta(k,e)$ and $\zeta(k,e)$ are respectively the average number density, Fourier transforms of $\eta(\bv{r},e)$ and
$\zeta(\bv{r},e)$. 
It is known that $\nu^*_1(k,e)$ and $\nu^*_2(k,e)$
are respectively given by 
\begin{equation}\label{nu1}
\nu^*_1(k,1) = 2(1-j_0(k)+2j_2(k))/(3k^2) ,
\end{equation}
and 
\begin{equation}\label{nu2}
\nu^*_2(k,1) = 2(1-j_0(k)-j_2(k))/(3k^2)
\end{equation}
for elastic hard spherical particles, 
where $j_l(k)$ with $l=0$ or $2$ is the $l$-th. order spherical Bessel function \cite{Kirkpatrick85,Kirkpatrick86,Das90,Marchetti,deSchepper,Alley83}.
Although we do not know how $\nu^*_1(k,e)$ and $\nu^*_2(k,e)$ depend on $e$, the explicit $e$-dependences
of $\nu^*_1(k,e)$ and $\nu^*_2(k,e)$ are not
important in this paper. Therefore, we are keeping discussion without their explicit forms.

It should be noted that sheared corrections to the structure factor or the pair-correlation function 
can be obtained within this theoretical
framework, though the equilibrium or the unsheared structure factor $S_0(k,e)\equiv S(k,e,\dot\gamma=0)$ should be determined by another method.
We adopt an approximate expression of the pair-correlation function 
for unsheared granular liquids obtained by Lutsko \cite{lutsko01},
which covers the equilibrium pair-correlation in the elastic limit (see Appendix A).  

This set of equations \eqref{mass}-\eqref{noise-corr} is a reasonable starting point for isothermal molecular liquids ($e=1$), 
once we use appropriate generalized
transport coefficients and $S_0(k,e=1)$.
However, the validity of this set of equations in describing isothermal sheared granular fluids might be controversial. 
Indeed, nobody has used the generalized fluctuating hydrodynamics for granular liquids, because we know that fluctuations cannot
be characterized by Gaussian. In addition, there are no explicit characteristics 
of granular liquids in the set of equations except for $e$-dependence of $S(k,e,\dot\gamma)$ and 
the generalized transport coefficients. 
Even when we use the hydrodynamic equations for granular liquids,
we should consider an equation for the granular temperature. 

Let us answer the above critical points to validate  the generalized fluctuating hydrodynamics in eqs. \eqref{mass}-\eqref{noise-corr}.
First, the granular temperature $T$ is not a true hydrodynamic variable
 but a relatively fast  variable because of the collisional energy loss in granular systems. 
Thus, we expect the fast relaxation to
a steady state of the temperature for sheared granular liquids,
which satisfies $T\propto \dot{\gamma}^2 / (1-e^2)$.
 Second, an isothermal situation is easily realized by the balance between the viscous heating  and  the collisional energy loss. 
In particular, it is known that USF is stable for small and nearly elastic systems  under Lees-Edwards
boundary condition. Even in physical situations, 
the heat conduction is not important in the bulk region far from the boundary.   
In these situations, we may assume that sheared granular liquids are nearly isothermal.
We also indicate that our previous studies clarify the formal similarities between sheared granular liquids and
sheared molecular liquids at a constant temperature \cite{Otsuki,hayakawa08}.
Through previous studies, we have recognized that the most important issue is to determine $S(k,e,\dot\gamma)$ 
for granular liquids \cite{hayakawa08},
which can be determined within the framework of the generalized fluctuating hydrodynamics.
Third, the fluctuating hydrodynamics has been used in describing
granular hydrodynamics for freely cooling cases \cite{Noije}.
Thus, we believe that the set of equations \eqref{mass}-\eqref{noise-corr} can be used even for sheared granular liquids.
Although there are uncovered regions of our approach 
in the description of
granular liquids, we expect that our approach can capture some aspects of sheared granular liquids. 
The validity of the model will be tested from the comparison between our theoretical prediction  
and the direct simulation of granular assemblies.

\section{Linearized equations around uniform shear flow and their solutions}

In this section, we analyze a set of the linearized equations around USF. 
As mentioned in the previous section,
we assume that USF is stable.  Thus, we only need to solve the linearized equations. This section consists of two
parts. In the first part, we
summarize the expression of the  linearized equations. In the second part, we explicitly write the solutions
of the linearized equations.

\subsection{Linearized equations}

Let us introduce the fluctuations of the hydrodynamic fields
$\delta n(\bv{r},t) \equiv n(\bv{r},t) - n_0$,
  $\delta \bv{u}(\bv{r},t) \equiv \bv{u}(\bv{r},t) - \bv{c}(\bv{r},t)$ with
$c_{\alpha}(\bv{r}) = \dot{\gamma} y \delta_{\alpha,x}$, and 
  the non-dimensionalized vector $\bv{z}(\bv{r},t)$, whose  Fourier transform is 
  given by
\begin{equation}
\hat{\bv{z}}^T({\bv{q}},t)= (\delta n(\bv{q},t), 
 \delta u_x(\bv{q},t)/(t_E^{-1} \sigma^4), \delta u_y(\bv{q},t)/(t_E^{-1} \sigma^4), \delta u_z(\bv{q},t)/(t_E^{-1} \sigma^4)). \label{z:def}
\end{equation}
Here, the Greek suffix $\alpha$ denotes the Cartesian component.
From eqs. (\ref{n:eq}), (\ref{u:eq}), and (\ref{z:def}), we obtain  the linearized evolution equation for $\tilde{\bv{z}}(\bv{k},\bar{t})\equiv \hat{\bv{z}}(\bv{q},t)$ as
\begin{equation}
\left(\partial_{\bar{t}} - \dot{\gamma}^* k_x \frac{\partial}{\partial k_y} \right) \tilde{\bv{z}}
+ {\sf L}\cdot \tilde{\bv{z}} = \tilde{\bv{R}},
 \label{Lin:eq}
 \end{equation}
 where 
 the time, the wavenumber and the shear rate have been non-dimensionalized  by
 $t= t_E \bar{t}$, $\bv{q} = \bv{k}/\sigma$,
 and $\dot{\gamma} = \dot{\gamma}^*/t_E$,
 respectively.  
Here, the matrix ${\sf L}$ is expanded as 
\begin{equation}
{\sf L} =  {\sf L}_{0} + \dot{\gamma}^* {\sf L}_{1} + \cdots, \label{matrix}
  \end{equation}
where ${\sf L}_{0}$ and ${\sf L}_{1}$ are respectively given by
\begin{eqnarray}
{\sf L}_{0} & = & 
\left[ 
\begin{array}{cccc}
0 & n_0 \sigma^3 ik_x & n_0 \sigma^3 ik_y & n_0 \sigma^3 ik_z \\
p^* ik_x  & (\nu^*_1-\nu^*_2)k_x^2 + \nu^*_2k^2 &
 (\nu^*_1-\nu^*_2)k_xk_y & (\nu^*_1-\nu^*_2)k_x k_z  \\
p^* ik_y & (\nu^*_1-\nu^*_2)k_y k_x & (\nu^*_1-\nu^*_2)k_y^2 + \nu^*_2 k^2 & 
(\nu^*_1-\nu^*_2)k_y k_z  \\
p^* ik_z & (\nu^*_1-\nu^*_2)k_z k_x & (\nu^*_1-\nu^*_2)k_z k_y & 
(\nu^*_1-\nu^*_2)k_z^2 + \nu^*_2 k^2 \\
\end{array} 
\right],
\end{eqnarray}
\begin{eqnarray}
{\sf L}_{1} & = & 
\left[ 
\begin{array}{cccc}
0 & 0 & 0 & 0  \\
0  & 0 & 1 & 0  \\
0  & 0  & 0 & 0  \\
0  & 0 & 0 & 0  \\
\end{array} 
\right]
\end{eqnarray}
with $p^*_1\equiv p^*(k,e) = A S_0(k,e)^{-1}$ 
and $A \equiv T /(mn_0 \sigma^5 t_E^{-2} )$.
The random vector $\tilde{\bv{R}}$  has four components 
\begin{eqnarray}\label{R_a}
\tilde{R}_1 & = & 0, \nonumber \\
\tilde{R}_{\alpha+1} & = & (mn_0 \sigma^4 t_E^{-2})^{-1} i \sigma k_\beta \Sigma_{\alpha\beta}^R(\bv{k},\bar{t}),
\end{eqnarray}
where $\alpha=1,2,3$ respectively correspond to $x, y$ and $z$.
Although $\dot\gamma\*$ dependence of $S(k,e,\dot\gamma)$ should appear in ${\sf L}_1$,
we simply ignore such terms. The validity of this simplification will be checked from the comparison of the results with our simulation.

\subsection{Solution of the linearized equations}

It is straightforward
to solve eq. \eqref{Lin:eq}. As mentioned in ref. \cite{Lutsko85},
its calculation can be simplified if we introduce the transformation from the Cartesian coordinate
to the oblique coordinate to decompose the longitudinal modes and the transverse mode.
However, the obtained results in the oblique coordinate are rather confusing, because of too many suffices.
In addition, the results are basically the same as those obtained by Lutsko and Dufty \cite{Lutsko85} with 
replacing the transport coefficients by the generalized transport coefficients,
and ignoring terms related with the fluctuation of the temperature.
In this paper, thus, we only  present the final results in the Cartesian coordinate.
The transformation between the Cartesian coordinate and the oblique coordinate is explained in Appendix B.

The solution of eq. (\ref{Lin:eq}) can be formally represented by  
\begin{equation}
\tilde{\bv{z}}(\bv{k},\bar{t}) = \sum_{j=1}^4 \int _{-\infty}^{\bar{t}} ds
\tilde{\bv{\psi}}^{(j)}(\bv{k},\bar{t}-s)
F^{(j)}(\tilde{\bv{k}}(\dot{\gamma}^*(s-\bar{t})),s), \label{z:sol}
\end{equation}
where 
\begin{eqnarray}
\tilde{\bv{\psi}}^{(j)}(\bv{k},\bar{t})
&\equiv&
\bv{\psi}^{(j)}(\bv{k}) \exp[  - \int_0^{\bar{t}} d\bar{\tau} \lambda^{(j)}( \tilde{\bv{k}}(\dot{\gamma}^*\bar{\tau}))] ,
\\
F^{(j)}(\bv{k},\bar{t}) &\equiv& \bv{\varphi}^{(j)}(\bv{k})\cdot\tilde{\bv{R}}(\bv{k},\bar{t})
\label{F^j}
\end{eqnarray}
with $\tilde{\bv{k}}(\bar{\tau}) \equiv (k_x,k_y + \bar{\tau} k_x,k_z)$.
Here, we have introduced
the right eigenvectors 
$\bv{\psi}^{(j)}(\bv{k})$, 
the associated biorthogonal vectors, {\it i.e.} the left eigenvectors $\bv{\varphi}^{(j)}(\bv{k})$,
and the eigenvalues $\lambda^{(j)}(\bv{k})$ satisfying
\begin{equation}
\left(- {\sf 1}\dot{\gamma}^* k_x \frac{\partial}{\partial k_y}  + {\sf L} \right)\cdot
\bv{\psi}^{(j)}(\bv{k}) = \lambda^{(j)}(\bv{k}) \bv{\psi}^{(j)}(\bv{k}).
\label{eigen:eq}
\end{equation}
We also note that $\bv{\psi}^{(j)}(\bv{k})$ and $\bv{\varphi}^{(j)}(\bv{k})$ satisfy
$\bv{\psi}^{(i)}(\bv{k})\cdot \bv{\varphi}^{(j)}(\bv{k})= \delta_{ij}$.

In order to obtain 
$\bv{\psi}^{(j)}(\bv{k})$, 
$\bv{\varphi}^{(j)}(\bv{k})$, 
and $\lambda^{(j)}(\bv{k})$,
we use the expansions
\begin{eqnarray}
\bv{\psi}^{(j)} & = & \bv{\psi}^{(j)}_{0} + \dot{\gamma}^* \bv{\psi}^{(j)}_{1} + \cdots, \label{expand1}\\
\bv{\varphi}^{(j)} & = &\bv{\varphi}^{(j)}_{0} + \dot{\gamma}^* \bv{\varphi}^{(j)}_{1} + \cdots, \label{expand2}\\
\lambda^{(j)} & = & \lambda^{(j)}_0 + \dot{\gamma}^* \lambda^{(j)}_1 + \cdots,\label{expand3}
\end{eqnarray}  
in terms of $\dot{\gamma}^*$.

We should note that the perturbation in terms of $\dot\gamma^*$ is not the expansion from an unsheared state
of granular liquids. Indeed, it is well-known that properties of sheared granular liquids completely differ
from those of freely cooling granular liquids. In the case of sheared granular liquids,
we obtain the relation
$\dot{\gamma}^* \sim \dot{\gamma} / \sqrt{T}
\sim \sqrt{1-e^2}$
from the balance between the viscous heating and the collisional energy loss. 
Thus, the expansion in terms of $\dot\gamma^*$ can be regarded as that 
by small inelasticity \cite{Alam08}.

Substituting eqs. (\ref{expand1}), (\ref{expand2}) and (\ref{expand3})
into eq. (\ref{eigen:eq}),
we obtain the zeroth and the first order perturbations as
\begin{eqnarray}
( {\sf L}_{0} - \lambda^{(j)}_0 {\sf 1})\cdot
\bv{\psi}^{(j)}_{0}(\bv{k}) & = & 0, \\
( {\sf L}_{(0)} - \lambda^{(j)}_0 {\sf 1})\cdot
\bv{\psi}^{(j)}_{1}(\bv{k}) +
\left( -  {\sf 1}k_x \frac{\partial}{\partial k_y}  + {\sf L}_{1} - \lambda^{(j)}_1 {\sf 1} \right)\cdot
\bv{\psi}^{(j)}_{0}(\bv{k}) 
& = & 0.
\end{eqnarray}
Solving these equations, we obtain the eigenvalues
\begin{eqnarray}
\lambda^{(1)} & = & \lambda_+ + \dot{\gamma}^* \frac{k_xk_y}{k^2} \xi^{(1)}(k), \label{lam1} \\
\lambda^{(2)} & = & \lambda_- + \dot{\gamma}^* \frac{k_xk_y}{k^2} \xi^{(2)}(k), 
\label{lam2}\\
\lambda^{(3)} & = & \nu^*_2(k,e)k^2 - \dot{\gamma}^* \frac{k_xk_y}{k^2}, \\
\lambda^{(4)} & = & \nu^*_2(k,e)k^2
\end{eqnarray}
within the approximation up to $O(\dot\gamma^*)$, where we have introduced
\begin{eqnarray}
\lambda_+ & = &\frac{\nu^*_1(k,e)k^2 + \sqrt{(\nu^*_1(k,e)k^2)^2 - 4n_0 \sigma^3p^*(k,e)k^2}}{2}, \\
\lambda_- & = &\frac{\nu^*_1(k,e)k^2 - \sqrt{(\nu^*_1(k,e)k^2)^2 - 4n_0 \sigma^3p^*(k,e)k^2}}{2},
\end{eqnarray}
\begin{equation} 
\xi^{(1)}(k)\equiv \frac{\lambda_+^2}{N_+^2} + \frac{n_0\sigma^3k^2}{2N_+^2}
k \partial_k p^*(k,e), \qquad
\xi^{(2)}(k)\equiv \frac{\lambda_-^2}{N_-^2} + \frac{n_0\sigma^3k^2}{2N_-^2}
k \partial_k p^*(k,e),
\end{equation}
and
\begin{equation}
N_+^2 = - n_0 \sigma^3p^*(k,e)k^2 + \lambda_+^2, \qquad
N_-^2 = - n_0 \sigma^3p^*(k,e)k^2 + \lambda_-^2.
\end{equation}
Similarly,
we obtain
the right eigenvectors
\begin{eqnarray}
\bv{\psi}^{(1)T} & = & \frac{1}{N_+}\left(ikn_0 \sigma^3,\lambda_+ \frac{k_x}{k},\lambda_+ \frac{k_y}{k},\lambda_+ \frac{k_z}{k} \right), 
\label{psi1}\\
\bv{\psi}^{(2)T} & = & \frac{1}{N_-}\left(ikn_0 \sigma^3,\lambda_- \frac{k_x}{k},\lambda_- \frac{k_y}{k},\lambda_- \frac{k_z}{k} \right), \label{psi2}\\
\bv{\psi}^{(3)} & = & \bv{\Psi}^{(3)} + M(\bv{k})\bv{\Psi}^{(4)}, \label{psi3}\\
\bv{\psi}^{(4)} & = & \bv{\Psi}^{(4)}, \\
\end{eqnarray}
and the left eigenvectors
\begin{eqnarray}
\label{phi^1}
\bv{\varphi}^{(1)} & = & \frac{1}{N_+}\left(ikp^*(k,e),\lambda_+ \frac{k_x}{k},\lambda_+ \frac{k_y}{k},\lambda_+ \frac{k_z}{k} \right), \label{phi1}\\
\bv{\varphi}^{(2)} & = & \frac{1}{N_-}\left(ikp^*(k,e),\lambda_- \frac{k_x}{k},\lambda_- \frac{k_y}{k},\lambda_- \frac{k_z}{k} \right), \label{phi2}\\
\bv{\varphi}^{(3)} & = & \bv{\Phi}^{(3)}, \\
\bv{\varphi}^{(4)} & = & -M(\bv{k})\bv{\Phi}^{(3)}+\bv{\Phi}^{(4)}, \label{phi^4}
\end{eqnarray}
where we have used 
\begin{eqnarray}
\bv{\Psi}^{(3)T} & = & \bv{\Phi}^{(3)} \equiv \left(0,-\frac{k_y k_x }{kk_\perp}, \frac{k_\perp}{k},-\frac{k_y k_x }{kk_\perp} \right) , \\
\bv{\Psi}^{(4)T} & = & \bv{\Phi}^{(4)}\equiv  \left(0, \frac{k_z}{k_\perp},0,-\frac{k_x}{k_\perp}\right),
\end{eqnarray}
and
\begin{equation}
M(\bv{k}) = - \frac{kk_z}{k_xk_\perp} \tan^{-1}(k_y/k_\perp) 
\end{equation}
with $k_\perp \equiv k^2 - k_y^2$.

\section{Correlation functions}

This section is the main part of this paper, in which we present the explicit forms of correlation functions.
This section consists of two parts. The first part summarizes the general results of correlation functions.
In the second part, we evaluate integrals included in correlation functions  to extract
the long-range behaviors of correlations.

\subsection{General results for correlation functions}

Let us introduce the correlation functions $\tilde{C}_{ii}(\bv{k},\bar{t})$
which satisfy
\begin{equation}
\left <  \tilde{z}_i(\bv{k},\bar{t}) \tilde{z}_i(\bv{k}',\bar{t})\right >= 
(2\pi)^3 \delta^3(\bv{k}+\bv{k}') \tilde{C}_{ii}(\bv{k},\bar{t}).
\end{equation}
Note that we do not use Einstein's sum rule for Latin subscripts.
Substituting eq. (\ref{z:sol}) into this equation,
we obtain
\begin{equation}
\tilde{C}_{ii}(\bv{k},\bar{t}) = 
\int_0^{\infty} d\bar{t} \sum_{l,m} 
\tilde{\psi}^{(l)}_i(\bv{k},\bar{t}) \tilde{\psi}^{(m)}_i(-\bv{k},\bar{t})
F^{(lm)}(\tilde{\bv{k}}(\dot{\gamma}^*\bar{t})), \label{ck:ex}
\end{equation}
where $\tilde{\psi}^{(l)}_i(\bv{k},\bar{t})$ is the $i$-th. component of $\tilde{\bv{\psi}}^{(l)}(\bv{k},\bar{t})$,
and $F^{(lm)}(\bv{k})$ satisfies
\begin{equation}
\left < F^{(l)}(\bv{k},\bar{t}) F^{(m)}(\bv{k}',\bar{t}')\right > = 
(2\pi)^3 \delta^3(\bv{k}+\bv{k}') \delta(\bar{t}-\bar{t}') F^{(lm)}(\bv{k}).
\end{equation}

From eqs. \eqref{R_a}, \eqref{F^j}  and \eqref{phi^1}-\eqref{phi^4}, 
the explicit forms of $F^{(lm)}(\bv{k})$ are given by
\begin{eqnarray}
F^{(11)}(\bv{k}) & = & - 2A k^2\nu^*_1(k,e) \frac{\lambda_+^2}{N_+^2},
\nonumber \\
F^{(22)}(\bv{k}) & = & - 2A k^2\nu^*_1(k,e) \frac{\lambda_-^2}{N_-^2},
\nonumber \\
F^{(12)}(\bv{k}) & = & F^{(21)}(\bv{k},\bar{t}) = - 2A k^2\nu^*_1(k,e) \frac{\lambda_+ \lambda_-}{N_+ N_-},
\nonumber \\
F^{(33)}(\bv{k}) & = &  2A k^2\nu^*_2(k,e),
\nonumber \\
F^{(44)}(\bv{k}) & = &  -(M(\bv{k})^2 + 1)F^{(33)}(\bv{k}),
\nonumber \\
F^{(34)}(\bv{k}) & = &  -F^{(43)}(\bv{k}) =  M(\bv{k}) F^{(33)}(\bv{k}).
\label{F_ij}
\end{eqnarray}
Thus, we can calculate any spatial correlation functions.

Let us explicitly write the spatial density correlation 
\begin{equation}
C_{nn}(\bv{r},\bar{t}) \equiv  \left < \delta 
n(\bv{r}+\bv{r}',\bar{t})
\delta n(\bv{r}',\bar{t}) \right >=\sigma^{-6} C_{11}(\bv{r},\bar{t})
\end{equation}
and the spatial momentum correlation 
\begin{equation}
C_{pp}(\bv{r},\bar{t}) \equiv \left < \bv{p}(\bv{r}+\bv{r}',\bar{t})
\cdot \bv{p}(\bv{r}',\bar{t}) \right >
\simeq (mn_0\sigma/t_E)^2 
\{ \tilde{C}_{22}(\bv{k},\bar{t}) +\tilde{C}_{33}(\bv{k},\bar{t})+\tilde{C}_{44}(\bv{k},\bar{t})\} \label{C_pp:def}
\end{equation}
with the momentum density $\bv{p}(\bv{r},\bar{t}) = mn(\bv{r},\bar{t}) \delta \bv{u}(\bv{r},\bar{t})$.
In eq. (\ref{C_pp:def}), we adopt the approximation
$\bv{p}(\bv{r},\bar{t}) \simeq mn_H \delta \bv{u}(\bv{r},\bar{t})$.
With the help of the inverse Fourier transform
\begin{eqnarray}
C_{nn}(\bv{r},\bar{t}) & = & \int \frac{d \bv{q}}{(2\pi)^3} \tilde{C}_{nn}(\bv{k},\bar{t}) e^{-i\bv{q} \cdot
\bv{r}}, \label{rev_n:four} \\
C_{pp}(\bv{r},\bar{t}) & = &  \int \frac{d \bv{q}}{(2\pi)^3} 
\tilde{C}_{pp}(\bv{k},\bar{t}) e^{-i\bv{q} \cdot
\bv{r}},  \label{rev_p:four}
\end{eqnarray}
and eq. (\ref{ck:ex}), 
we obtain the steady solutions (see Appendices C and D), 
\begin{eqnarray}
\tilde{C}_{nn}(\bv{k}) & = & n_0 \left \{
S_0(k,e) + \dot{\gamma}^* \tilde{\Delta}_1(\bv{k})
\right \},
\label{C_nn}
\\
 \tilde{C}_{pp}(\bv{k}) & = & (mn_0\sigma / t_E)^2 \{
 -\tilde{D}_2(\bv{k}) + \tilde{D}_3(\bv{k}) -\tilde{D}_4(\bv{k})\},
\label{C_pp}
 \end{eqnarray}
where  $\tilde{C}_{nn}(\bv{k})\equiv \lim_{\bar{t}\to\infty}\tilde{C}_{nn}(\bv{k},\bar{t})$, and
$S_0(k,e)$ is the structure factor for unsheared case.
$\tilde{D}_j(\bv{k})$ with $j=2,3,4$ in eq. (\ref{C_pp}) are given by
\begin{eqnarray}
\tilde{D}_{2}(\bv{k}) & = &
-A \left \{
1 + \dot{\gamma}^* \tilde{\Delta}_2(\bv{k})
\right \},  \label{D:2}\\
\tilde{D}_{3}(\bv{k}) & = &
A \left \{
1 + 2\dot{\gamma}^* \tilde{\Delta}_3(\bv{k})
\right \}, \label{D:3} \\
\tilde{D}_{4}(\bv{k}) & = &
-A \left \{
1 + 2\dot{\gamma}^* \tilde{\Delta}_4(\bv{k})
\right \}. \label{D:4}
\end{eqnarray}
The derivation of eqs. (\ref{C_nn})-(\ref{D:4}) and the 
explicit forms of $\tilde{\Delta}_j(\bv{k},\bar{t})$ with $j=1,2,3,4$ 
are presented in Appendix \ref{C_nn:app} and \ref{C_pp:app}.

From (\ref{rev_n:four})-(\ref{D:4}),
we finally obtain
\begin{eqnarray}
C_{nn}(\bv{r}) & = & n_0^2 \left \{
g_0(r,e) - 1 + \dot{\gamma}^* \Delta_{1}(\bv{r})
\right \} + n_0 \delta(\bv{r}), \label{C_nn:r} \\
C_{pp}(\bv{r}) & = & \frac{T_0}{mn_0}
\left [
3\delta(\bv{r}) + \sigma^{-3} \dot{\gamma}^* \left \{ \Delta_{2}(\bv{r}) + \Delta_{3}(\bv{r}) 
+\Delta_{4}(\bv{r}) \right \}
\right ],\label{C_pp:r}
\end{eqnarray}
where $g_0(r,e)$ is the pair-correlation function for unsheared cases, and
\begin{eqnarray}
\Delta_{\alpha}(\bv{r}) & = & \int \frac{d \bv{q}}
{(2\pi)^3} \tilde{\Delta}_{\alpha}(\bv{k}) e^{-i\bv{q} \cdot \bv{r}}.
\end{eqnarray}

The determination of $g_0(r,e)$ or $S_0(k,e)$ for $e<1$ is highly nontrivial. Indeed, we cannot keep
any homogeneous cooling state (HCS)  without artificial controls of the systems.
However, Lutsko \cite{lutsko01} obtained an approximate expression of $g_0(r,e)$ for HCS.
He also verified that his approximate expression works well from the comparison between the theory and the simulation of HCS. 
As stated in Section II,  we adopt his expression in this paper (Appendix A).
We also note that the structure of the liquids in eq. 
\eqref{C_nn:r} can be represented 
by the linear contribution of the homogeneous terms and the sheared term.

\subsection{Long-range correlation}

Let us  demonstrate the existence of the long-range correlation in
$C_{nn}(\bv{r})$ and $C_{pp}(\bv{r})$.
Let the angular average of any function $f(\bv{r})$ be denoted by
$\bar{f}(r) \equiv  \int d \Omega f(\bv{r})/ (4 \pi) $.
As shown in Appendix E, the asymptotic forms of $\bar{\Delta}_j(r)$  ($j=1,2,3$) satisfy
\begin{eqnarray}
\bar{\Delta}_1(r)  & \propto & r^{-11/3}, \qquad  r \gg l_c,  \label{Delta1:asym}\\
\bar{\Delta}_2(r)  & \propto & r^{-11/3}, \qquad  r \gg l_c,  \label{Delta2:asym}\\
\bar{\Delta}_3(r)  & \propto & r^{-5/3}, \qquad  r \gg l_c,\label{Delta3:asym}
 \\
\bar{\Delta}_4(r)  & \propto & r^{-5/3}, \qquad  r \gg l_c,\label{Delta4:asym}
\end{eqnarray} 
where $l_c \equiv \sigma / \sqrt{\dot{\gamma}^*}$. 
Substituting these results into \eqref{C_nn:r} and \eqref{C_pp:r}
the long-range parts of $\bar{C}_{nn}(r)$
and $\bar{C}_{pp}(r)$
respectively satisfy
\begin{eqnarray}\label{tail-Cnn}
\bar{C}_{nn}(r)  & \propto & \left ( \frac{r}{l_c} \right )^{-11/3}, \qquad r \gg l_c,\\
\bar{C}_{pp}(r)  & \propto & \left ( \frac{r}{l_c} \right )^{-5/3}, \qquad r \gg l_c. 
\label{tail-Cpp}\end{eqnarray}
This long-range correlation $C_{pp}(\bv{r})$ is known for isothermal sheared elastic fluids \cite{Lutsko85,Lutsko02,Wada},
and has been verified in sheared dilute granular fluids \cite{otsuki08a}.

\section{Comparison between theory and simulation}


To verify the validity of our theoretical prediction, 
we perform the event-driven molecular dynamics simulation for  three-dimensional hard spheres. 
In our simulation the time scale is mesured by
$\tau_0\equiv \sigma\sqrt{m/T_0}$ where $T_0$ is the averaged
initial temperature.
Particles are confined in a cell under the Lees-Edwards boundary condition, in which each linear dimension is 
$L$. The number of  particles is not fixed in our simulation,
but we control the system size $L$ and the volume fraction $\phi$ as well as the restitution constant $e$.
The initial state at the time $t=0$ is the equilibrium state,
and we will show the correlation functions at $t=20 \tau_0$ for $\phi=0.50$ and $0.37$, 
and $t=40 \tau_0$ for $\phi=0.185$ as $\bar{C}_{nn}(r)$ and 
$\bar{C}_{pp}(r)$, where the system is considered in a steady state.
We also choose the shear rate to keep the steady temperature unity in the dimensionless unit, except for the data of Fig. \ref{cnn_ep0.83_A4.0},
where the temperature in the steady state is $1.0 T_0$ for $\dot \gamma = 0.92 \tau_0^{-1}$ or $4.0T_0$ for
$\dot \gamma = 1.84 \tau_0^{-1}$.

\begin{figure}
  \includegraphics[height=.3\textheight]{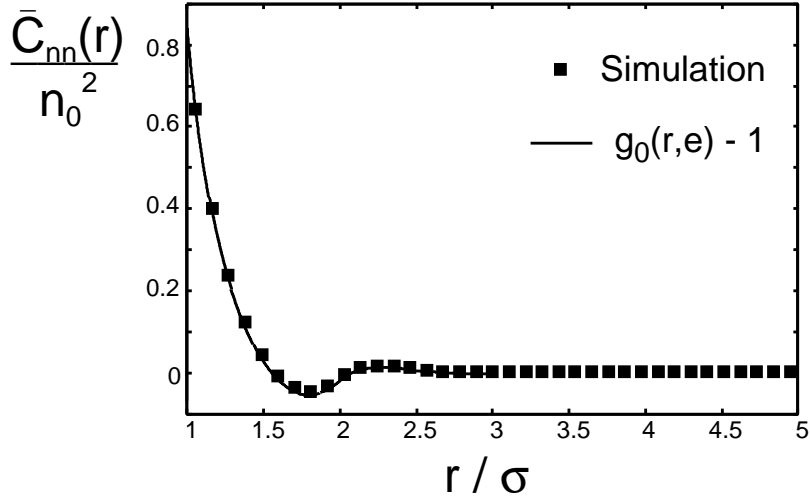}
  \caption{The density correlation function $\bar{C}_{nn}(r)$
for the volume fraction $\phi= 0.185$ with  $L = 89\sigma$  and $e=0.83$
as a function of the distance $r$.
The solid line represents $g_0(r,e)$ obtained by Lutsko\cite{lutsko01}. 
}
  \label{gr_ep0.83_A4.0}
\end{figure}

\begin{figure}
  \includegraphics[height=.3\textheight]{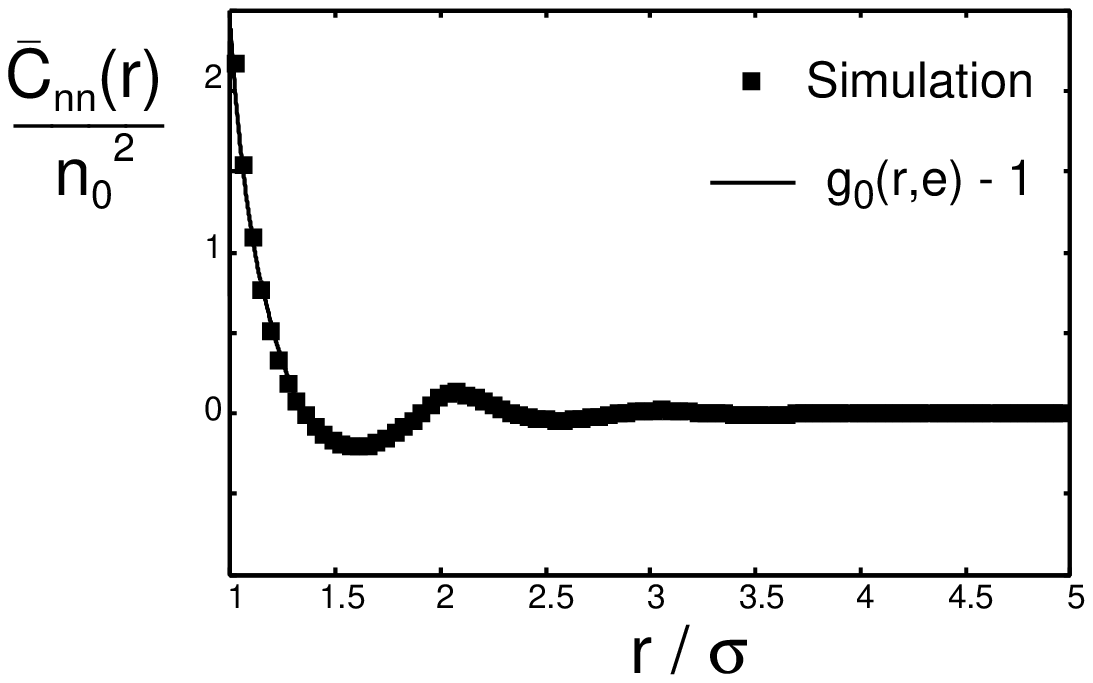}
  \caption{The density correlation function $\bar{C}_{nn}(r)$
for the volume fraction $\phi= 0.37$ with  $L = 72\sigma$  and $e=0.90$
as a function of the distance $r$.
The solid line expresses $g_0(r,e)$ obtained by Lutsko\cite{lutsko01}.
}
  \label{gr_ep0.9_A2.0}
\end{figure}

\begin{figure}
  \includegraphics[height=.3\textheight]{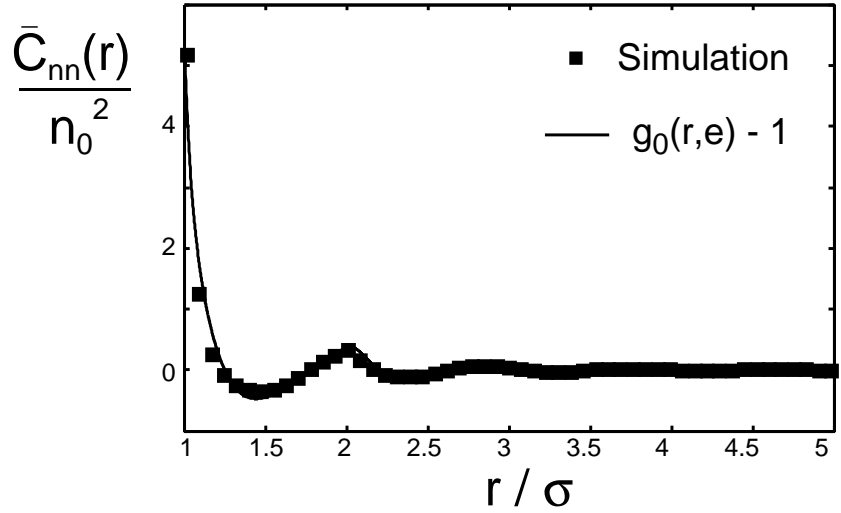}
  \caption{The density correlation function $\bar{C}_{nn}(r)$
for the volume fraction $\phi= 0.50$ with  $L = 32.5\sigma$  and $e=0.90$
as a function of the distance $r$.
The solid line expresses $g_0(r,e)$ obtained by Lutsko\cite{lutsko01}.
}
  \label{gr_ep0.9_nu0.5}
\end{figure}

Figures \ref{gr_ep0.83_A4.0}-\ref{gr_ep0.9_nu0.5}
show the behaviors of $\bar{C}_{nn}(r)$ for $r\le 5\sigma$ 
at $\phi = 0.185, 0.37$, and  $0.50$, respectively.
The restitution coefficient $e$ is $0.83$ for  
Fig. \ref{gr_ep0.83_A4.0},  and
$e$ is $0.90$ for  Figs. \ref{gr_ep0.9_A2.0} and \ref{gr_ep0.9_nu0.5}.
The solid line represents $g_0(r,e)$ obtained by Lutsko \cite{lutsko01} without any fitting parameters. 
Although we omit contributions of shear rate to
$\bar{C}_{nn}(r)$ because of its simplicity,
the agreement between the results of our simulation and our theory seems
to be perfect.
Thus, it is hard to find
 any contributions of the shear in the short-range structure of the density correlation function, as is known
in dense elastic liquids.

However, the above results do not mean that contributions of the shear 
to the density correlation function are not important.
Indeed, we find the existence of a power law tail
which might be consistent with the theoretical prediction  $\bar{C}_{nn}(r)\sim r^{-11/3}$ in  eq. \eqref{tail-Cnn} for $r\gg \sigma$ 
(see Fig. \ref{cnn_ep0.83_A4.0} for 
$\phi = 0.185$, and $\dot \gamma = 0.92\tau_0^{-1}$ and  $1.84\tau_0^{-1}$).
It should be noted that 
the shear rate dependence of $\bar{C}_{nn}(r)$ cannot be observed, because 
the scaled shear rate $\dot \gamma^* \sim \dot \gamma / \sqrt{T} \sim
\sqrt{1-e^2}$ in eq. \eqref{C_nn:r} is independent of  $\dot \gamma$ in the steady state.
The range of the tail obeying a power law is not wide enough to verify the theoretical prediction 
in Fig.  \ref{cnn_ep0.83_A4.0} because of the large statistical errors.
However, as will be shown in Figs. \ref{cuu_ep0.9_A2.0} and 8,
the numerical data for  $\bar{C}_{pp}(r)$ are consistent with the theoretical prediction.
To confirm the quantitative accuracy of our theory  in $\bar{C}_{nn}(r)$, we need more 
extensive simulations to reduce the statistical errors.

\begin{figure}
  \includegraphics[height=.3\textheight]{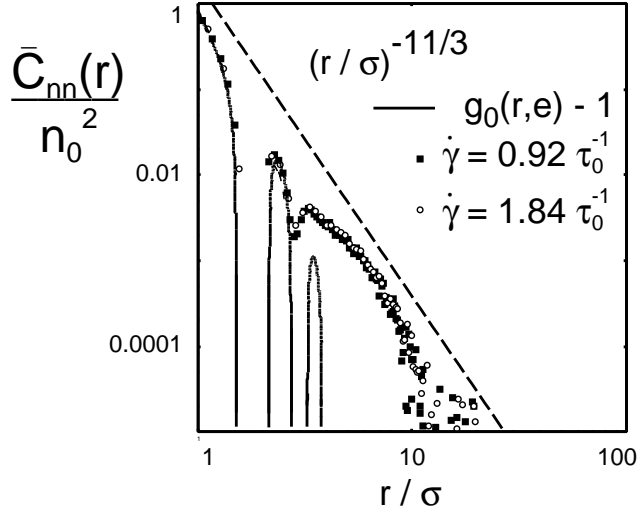}
  \caption{The double-log plot of $\bar{C}_{nn}(r)$
for the volume fraction $\phi =0.185$ with   $L = 89\sigma$, 
and $e=0.83$ as a function of the distance $r$. Here, the solid line is that for homogeneous case obtained by
Lutsko\cite{lutsko01}, and the plotted data are obtained in the case of $\dot\gamma=0.92\tau_0^{-1}$ and $1.84\tau_0^{-1}$.}
  \label{cnn_ep0.83_A4.0}
\end{figure}

\begin{figure}
  \includegraphics[height=.3\textheight]{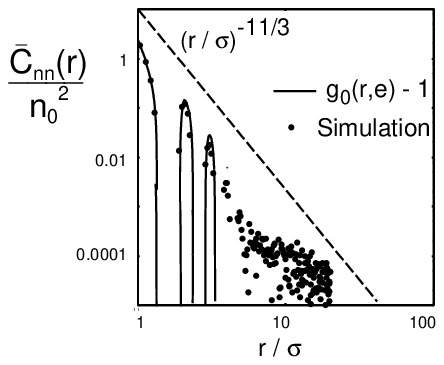}
  \caption{The double-log plot of $\bar{C}_{nn}(r)$
for the volume fraction $\phi =0.37$ with   $L = 72\sigma$, 
$e=0.90$ as a function of the distance $r$. Here, the solid line is that for homogeneous case obtained by
Lutsko.\cite{lutsko01}}
  \label{cnn_ep0.9_A2.0}
\end{figure}

\begin{figure}
  \includegraphics[height=.3\textheight]{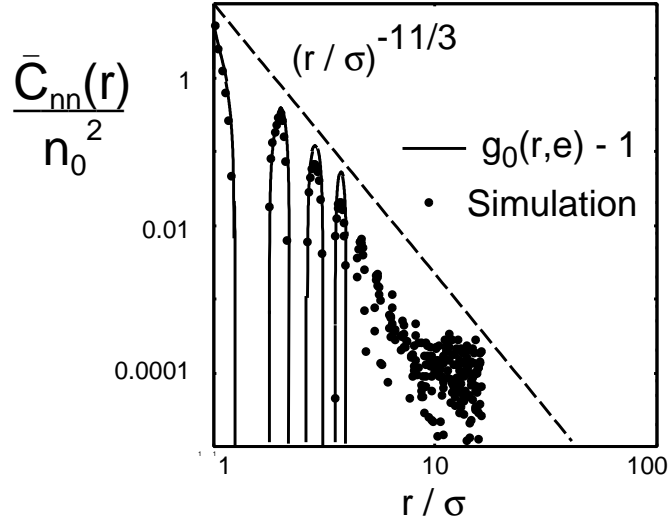}
  \caption{The double-log plot of  $\bar{C}_{nn}(r)$
for the volume fraction $\phi =0.50$ with   $L = 32.5\sigma$, 
$e=0.90$ as a function of the distance $r$. Here, the solid line is that for homogeneous case obtained by
Lutsko.\cite{lutsko01}}
  \label{cnn_ep0.9_nu0.5}
\end{figure}

Figures \ref{cnn_ep0.9_A2.0} and \ref{cnn_ep0.9_nu0.5} shows 
the behavior of $\bar{C}_{nn}(r)$ 
for 
$\phi = 0.37$ and  $0.50$, respectively.
We use $e=0.90$ and $L=72 \sigma$ for $\phi=0.37$, and $e=0.90$ and $L=32.5 \sigma$ for $\phi=0.50$.
Although we can identify the existence of a power law tail, it is hard to verify  whether the tail satisfies 
$r^{-11/3}$ for $\phi = 0.37$ and  $0.50$
because of  the large statistical errors.

The long-range momentum correlation function 
which satisfies $r^{-5/3}$ can be observed more clearly than the case of the density correlation function as shown in
 Figs. \ref{cuu_ep0.9_A2.0} and \ref{cuu_ep0.9_nu0.5}.
 Figure \ref{cuu_ep0.9_A2.0} is  the result of our simulation for 
$\phi=0.37$ and $e=0.90$. 
 Figure \ref{cuu_ep0.9_nu0.5} shows  the result for 
$\phi=0.50$ and $e=0.90$. 
Although there is an apparent finite size effect,
$\bar{C}_{pp}(r)$ can be an universal function of $r/L$, and  
decays faster than power-law function for $r > 0.4L$
 (see Figs. 7 and 8).
We will have to check whether the oscillation of $\bar{C}_{pp}(r)$ in small $r$ can be understood by our theory. 
To obtain the complete forms of $\bar{C}_{nn}(r)$ and $\bar{C}_{pp}(r)$
we need to solve the eigenvalue problem of inelastic Enskog operator
and determine $\nu_1^*(k,e)$ and $\nu_2^*(k,e)$ explicitly.
This will be our future work.

\begin{figure}
  \includegraphics[height=.3\textheight]{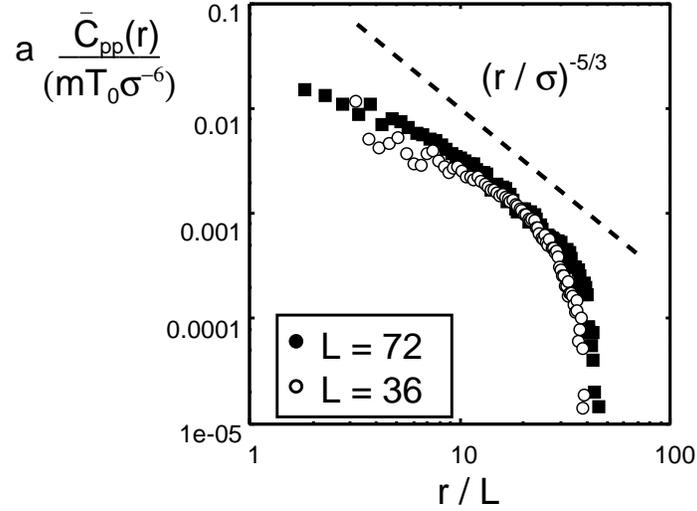}
  \caption{The double-log plot of  $\bar{C}_{pp}(r)$
for the volume fraction $\phi= 0.37$  with  $L/\sigma = 36, 72$ and $e=0.90$ as a function of the distance $r$. 
Here, $a$ is a fitting parameter, which is $0.4$ for $L=36\sigma$ or   $1.0$ 
  for $L=72\sigma$.}
  \label{cuu_ep0.9_A2.0}
\end{figure}

\begin{figure}
  \includegraphics[height=.3\textheight]{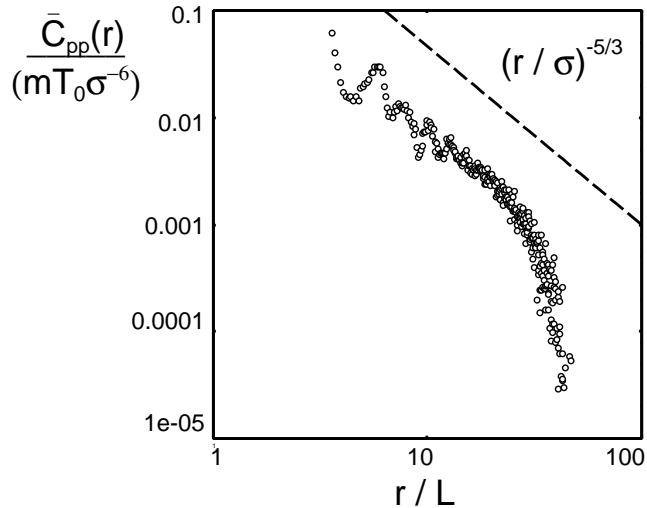}
  \caption{The double-log plot of  $\bar{C}_{pp}(r)$
for the volume fraction $\phi= 0.50$  with  $L = 32.5\sigma$ and $e=0.90$ as a function of the distance $r$.}
  \label{cuu_ep0.9_nu0.5}
\end{figure}

\begin{figure}
  \includegraphics[height=.3\textheight]{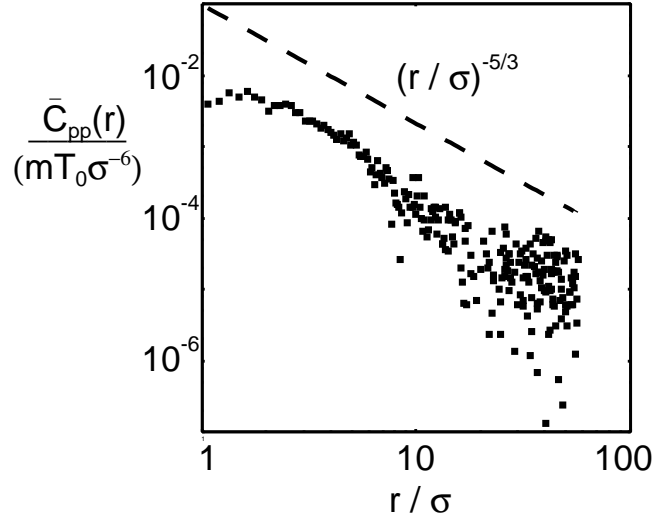}
  \caption{The double-log plot of   $\bar{C}_{pp}(r)$
for the volume fraction $\phi= 0.093$  with  $L = 112\sigma$  
and $e=1.0$ at $t=40 \tau_0$ as a function of the distance $r$.}
  \label{cuu_ep1.0_A8.0}
\end{figure}

\begin{figure}
  \includegraphics[height=.3\textheight]{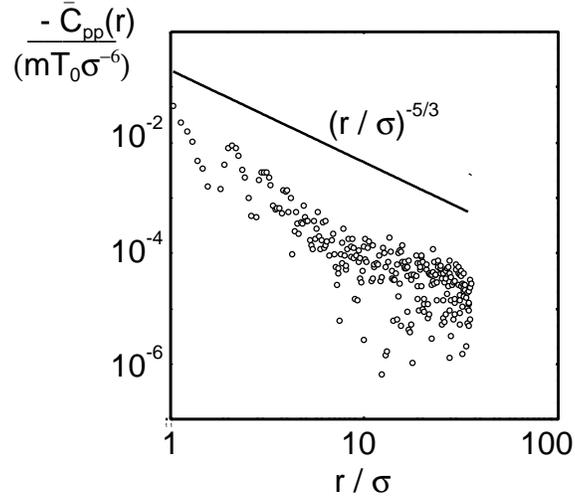}
  \caption{The double-log plot of   $\bar{C}_{pp}(r)$
for the volume fraction $\phi= 0.37$  with  $L = 72\sigma$  
and $e=1.0$ at $t=30 \tau_0$ as a function of the distance $r$.}
  \label{cuu_ep1.0_A2.0}
\end{figure}

In this paper, we mainly focus on the results of sheared granular liquids, because granular liquids are more non-trivial than sheared isothermal elastic liquids,
in which particles collide with each other without any loss of energy. We also perform the molecular dynamics simulation for sheared elastic liquids 
with the velocity scaling thermostat to keep  a constant temperature. Figures \ref{cuu_ep1.0_A8.0} and \ref{cuu_ep1.0_A2.0} are 
the results of the momentum correlation function
of sheared elastic liquids
with $\phi=0.093$ and $\phi=0.37$, respectively.
These results also support 
the power law $\bar{C}_{pp}(r)\sim r^{-5/3}$ even in the elastic case,
although $\bar{C}_{pp}(r)$ has the negative value for $\phi=0.37$
due to the back scattering effect. 
Although the existence of a power law $\bar{C}_{pp}(r)\sim r^{-5/3}$ has been believed \cite{Lutsko85},
it is the first time to verify the existence of such a tail from the molecular dynamics simulation.

\section{Discussion and Conclusion}

Now, let us discuss our result.
We should discuss the validity of the generalized fluctuating hydrodynamics. (i) Can the noise 
be Gaussian? The answer is no. However, the non-Gaussian nature may not be important in our framework, 
because we only discuss
the two-point correlation functions. 
When we discuss higher correlation functions, our theoretical treatment should be insufficient.
(ii) What are contributions of the energy fluctuations? 
It might be important to include effects of the temperature fluctuation for denser cases.
When we take into account such fluctuations, 
we must solve a cubic equation 
for the eigenvalues satisfying eq. (\ref{eigen:eq})
to obtain the explicit form of  $\tilde{C}_{ii}( \bv{k})$.
The analytic expression of the solution is, however,  rather complicated.
Hence, we do not take into account the fluctuations of the kinetic temperature.
This treatment may be justified as long as the USF is stable.
Indeed,  the temperature is immediately relaxed 
to be uniform.
(iii) The contribution of inelastic collisions also appears through
the inelastic Enskog operator for granular gases. 
At present, we have not solved the eigenvalue problem of the inelastic Enskog operator. In this sense,
our treatment in this paper is far from the complete. The complete treatment will be discussed elsewhere.

\begin{figure}
  \includegraphics[height=.3\textheight]{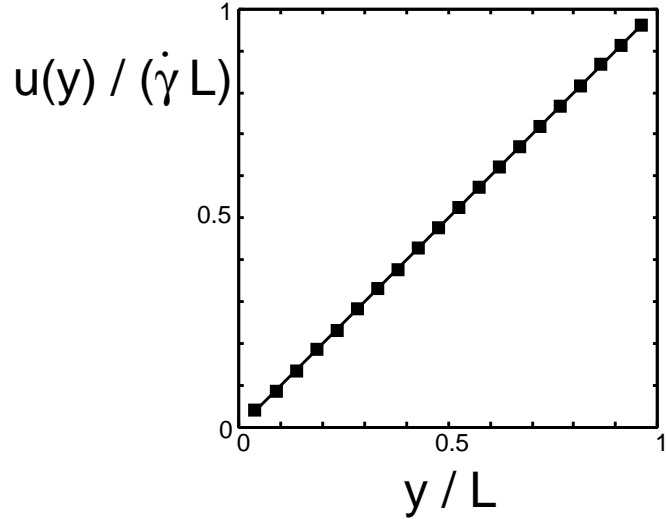}
  \caption{ The profile of the velocity $u(y)$ along the flow direction
  as a function of the position $y$ for $\phi= 0.37$,  $e=0.90$ and $L=32\sigma$
  with $t=20 \tau_0$.}
  \label{ugrad}
\end{figure}

\begin{figure}
  \includegraphics[height=.3\textheight]{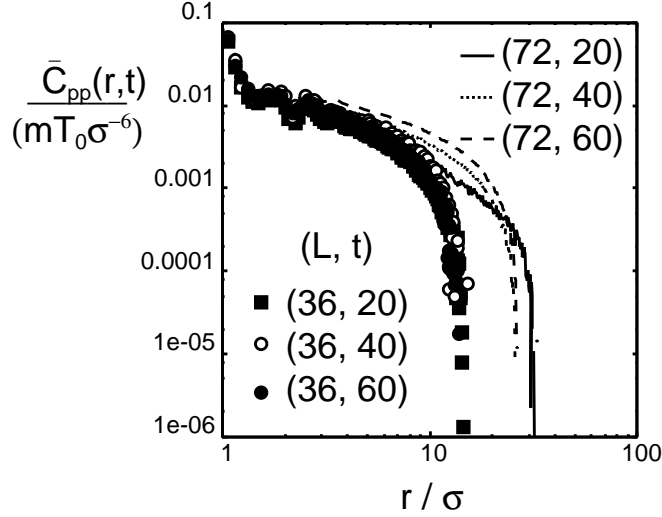}
  \caption{(Color on line) The time dependence of  
  $\bar{C}_{pp}(r)$
for the volume fraction $\phi= 0.37$  at $t = 20 \tau_0, 40 \tau_0, 60 \tau_0$ for $L = 36\sigma$ and  $72\sigma$ in the case of $e=0.90$.}
  \label{cuu_time_dep}
\end{figure}

Serious contributions of inelastic collisions appear in large systems, because  USF is unstable
for
larger and highly inelastic systems.  
It should be noted that the shear flow is induced by the
Lees-Edwards boundary condition.
Using this boundary condition,
USF is actually realized
as shown in  Fig. \ref{ugrad}, which is contrast to 
the shear flow obtained by moving a physical wall \cite{hisao07a}.
However, 
our long-time simulation for larger systems is deviated from the theoretical prediction.
In such a system, the linearized generalized hydrodynamics cannot be used, and the power law correlation function may
disappear.  This non-stationary tendency expected from unstable USF in larger systems can be observed in our simulation.
Figure \ref{cuu_time_dep} shows the time evolution of $\bar{C}_{pp}(r)$ for 
the system sizes $L=36\sigma$ and $L=72\sigma$ in the case of $e=0.90$ and $\phi=0.37$.
It is clear that the result of a smaller system ($L=36\sigma$) converges, but the result of a larger system ($L=72\sigma$) does not converge.
Although we have not identified the reason why we could not obtain the converged results in larger systems,
 it is likely to take place of the evolution of non-uniform structure \cite{hisao07a,alam1,alam2,Lee}. 

It should be noted that 
the instability of USF
can be reduced under the gravity, which is not included in this
paper \cite{silbert,namiko05}. 
Hence, we expect that the predicted long-range correlation can be
observed in experiments for sheared granular materials
\cite{kudrolli}, where the stable shear flow is realized.

In this paper, we demonstrate that the framework based on the generalized hydrodynamics is useful to determine
the structure factor or the pair-correlation function. This result is significant in the following two senses.
First, we clarify the significant contribution of the shear to the structure factor of dense isothermal liquids,
which is usually ignored in the papers to discuss glassy behaviors of sheared liquids \cite{cates1,miyazaki}. Second, the determination
of the structure factor gives complementary information to MCT for sheared granular liquids \cite{hayakawa08}.

In conclusion, we apply the generalized fluctuating hydrodynamics to isothermal sheared moderate dense liquids for
both elastic cases and inelastic cases. The theory predicts that the density correlation function and the momentum correlation
function, respectively, obey power laws  $r^{-11/3}$ and $r^{-5/3}$. Our theory
can be valid even for  the short-range scale for $\sigma\le r\le 5\sigma$.
The density correlation function for the short range scale can be approximated by the theory introduced by Lutsko \cite{lutsko01}.
These results have been verified through the comparison with the simulation.

\vspace*{0.5cm}

We thank H. Wada for fruitful discussions.
This work is partially supported by Ministry of Education, Culture, Science and Technology (MEXT), Japan (Grant No. 18540371) and the Grant-in-Aid for the global COE program
"The Next Generation of Physics, Spun from Universality and Emergence"
from the Ministry of Education, Culture, Sports, Science and
Technology (MEXT) of Japan.
One of the authors (M. O.) thanks the Yukawa Foundation for the financial
support.
The numerical calculations were carried out on Altix3700 BX2 at YITP in Kyoto University.

\appendix

\section{Approximate expression of $g_0(r,e)$}
\label{g_0:app}

In this Appendix, we show the outline of unsheared pair-correlation function obtained by Lutsko \cite{lutsko01},
who developed the generalized mean spherical approximation in describing a homogeneous cooling process of granular liquids.

The approximate expression of  $g_0(e,r)$ is obtained from
the inverse Laplace  transform of the function $G(t)$
satisfying 
\begin{eqnarray}
G(t) = \int_0^{\infty} dr e^{-tr} r g_0(r,e),
\end{eqnarray}
where $G(t)$ is given by
 \begin{eqnarray}
G(t) = \frac{t F(t) e^{-t}}{1+12 \phi F(t) e^{-t}}
\end{eqnarray}
with the volume fraction $\phi \equiv n_0 \sigma^3 \pi / 6$.
Here, we introduce $F(t)$ as
 \begin{eqnarray}
F(t) = \frac{-(1 + A_1 t + A_2 t^2)}{12 \phi (S_0 + S_1 t+ S_2 t^2 + S_3 t^3
+ S_4 t^4)},
\end{eqnarray}
where $S_0=1$, $S_1=A_1 - 1$, $S_2=A_2 - A_1 + 1/2$, $S_3=-A_2+A_1/2
- (1+2\phi)/(12\phi)$, $S_4=A_2/2 - (1+2\phi)/(12\phi) A_1
+(2+\phi)/(24\phi)$ with
 \begin{eqnarray}
A_1 & = &\frac{1}{2} + \sqrt{\frac{(\phi-1)^2 -  (6\phi g_0(\sigma,e) +1)Z}
{12\phi \{ (2+\phi) - 2g_0(\sigma,e) (\phi -1)^2 \} }}, \nonumber \\
A_2 & = & g_0(\sigma,e) \frac{(1+2\phi)A_1 - (2+\phi)/2}{1 + 6\phi g_0(\sigma,e)}.
\label{A:def}
\end{eqnarray}
In eq. (\ref{A:def}), the pair-correlation function $g_0(\sigma,e)$
at contact is given by the equilibrium pair-correlation function at
contact $g_{\rm eq}(\sigma) = (1-\phi/2)/(1-\phi)^3$
as $g_0(\sigma,e) = (1+e) g_{\rm eq}(\sigma) / (2e)$.
In addition, $Z$ is given as
 \begin{eqnarray}
Z = \left (  1 + \frac{1+e}{2} (Z_{\rm eq}^{-1} - 1 ) \right )^{-1},
\end{eqnarray}
where
 \begin{eqnarray}
Z_{\rm eq} = \left( \frac{\partial }{\partial n} \bar{P} \right )^{-1},
\end{eqnarray}
with 
$\bar{P} \equiv n(1+4\phi g_{\rm eq}(\sigma))/m$.

\section{Transformation between Cartesian coordinate and oblique coordinate}

In the calculation of eqs. (\ref{lam1})-(\ref{phi^4}), (\ref{F_ij}),
and (\ref{C_pp}), it is convenient to use the oblique coordinate,
where the vector $\hat{\bv{z}}$ of eq. (\ref{z:def}) 
in the Cartesian coordinate is transformed to the vector $\bar{\bv{z}}$
in the oblique coordinate as
 \begin{eqnarray}
\hat{\bv{z}}={\sf T} \bar{\bv{z}} ,
 \end{eqnarray}
where the matrix ${\sf T}$ is given by
\begin{eqnarray}
{\sf T} & = & 
\left[ 
\begin{array}{cccc}
1 & 0 & 0 &  \\
0  & k_x / k & -k_y k_x /(k k_\perp) & k_z / k_\perp  \\
0  & k_y / k &  k_\perp / k  & 0  \\
0  & k_z / k & -k_y k_z /(k k_\perp) & -k_x / k_\perp  \\
\end{array} 
\right].
\end{eqnarray}
Here, we note that the second, the third and the fourth components of $\bar{\bv{z}}$ are given by
$\bar{z}_2 = \delta u_\parallel$,
$\bar{z}_3 = \delta u_t$, and
$\bar{z}_4 = \delta u_s$,
where
$\delta u_\parallel = \bv{e}_\parallel \cdot \delta \bv{u}$,
$\delta u_t = \bv{e}_t \cdot \delta \bv{u}$,
$\delta u_s = \bv{e}_s \cdot \delta \bv{u}$,
and 
\begin{equation}
\bv{u} = \delta u_\parallel \bv{e}_\parallel + \delta u_t \bv{e}_t +
\delta u_s \bv{e}_s
\end{equation}
 with 
$\bv{e}_\parallel^{T} = ( k_x / k, k_y / k, k_z / k)$,
$\bv{e}_t^{T} = ( -k_y k_x /(k k_\perp), k_\perp/k, -k_y k_z /(k k_\perp))$,
and 
$\bv{e}_s^{T} = ( k_z /k_\perp, 0, -k_x / k_\perp)$.

From this transformation, the right eigenvectors and
the left eigenvectors in the oblique coordinate are respectively given by
\begin{eqnarray}
\bar{\bv{\psi}}^{(1)T} & = & \frac{1}{N_+}(ikn_0 \sigma^3,\lambda_+ ,0 , 0), \\
\bar{\bv{\psi}}^{(2)T} & = & \frac{1}{N_-}(ikn_0 \sigma^3,\lambda_- ,0 , 0), \\
\bar{\bv{\psi}}^{(3)T} & = & (0,0,1,M(\bv{k})), \\
\bar{\bv{\psi}}^{(4)T} & = & (0, 0, 0, 1), \\
\end{eqnarray}
and 
\begin{eqnarray}
\bar{\bv{\varphi}}^{(1)} & = & \frac{1}{N_+}(ikp^*(k,e),\lambda_+ , 0, 0),\\
\bar{\bv{\varphi}}^{(2)} & = & \frac{1}{N_-}(ikp^*(k,e),\lambda_- , 0, 0), \\
\bar{\bv{\varphi}}^{(3)} & = & (0,0,1,0), \\
\bar{\bv{\varphi}}^{(4)} & = & (0,0,-M(\bv{k}),1).\\
\end{eqnarray}

\section{Explicit representation of $\tilde{C}_{nn}(\bv{k})$}
\label{C_nn:app}

In this Appendix, 
we explicitly derive eq. (\ref{C_nn}).
$\tilde{C}_{nn}(\bv{k})$ is related to $\tilde{C}_{11}(\bv{k})$ as
$\tilde{C}_{nn}(\bv{k}) = \sigma^{-3} \tilde{C}_{11}(\bv{k})$.
Using eq. (\ref{ck:ex}),
we obtain
\begin{eqnarray}
\tilde{C}_{11}(\bv{k}) & = &
\int_0^{\infty} d\bar{t} \sum_{l,m} 
\tilde{\psi}^{(l)}_1(\bv{k},\bar{t}) \tilde{\psi}^{(m)}_1(-\bv{k},\bar{t})
F^{(lm)}(\tilde{\bv{k}}(\dot{\gamma}^*\bar{t})) \nonumber \\
& = &  
4An_0^2 \sigma^{6} \int_0^{\infty} d\bar{t} 
e^{-\int_0^{\bar{t}} d\bar{s}  \nu_1^*(\tilde{\bv{k}}(\dot{\gamma}^*\bar{s}))
\tilde{k}(\dot{\gamma}^*\bar{s})^2 } 
e^{-\int_0^{\bar{t}} d\bar{s} \dot{\gamma}^*
\frac{k_x\tilde{k}_y(\dot{\gamma}^*\bar{s})}
{\tilde{k}(\dot{\gamma}^*\bar{s})^2}
\{ 
  \xi^{(1)} (k(\dot{\gamma}^*\bar{s}) )
+ \xi^{(2)} (k(\dot{\gamma}^*\bar{s}) )
  \}} \nonumber \\
  & & \times
\frac{k^2}{N_+(k)N_-(k)}
\frac{\lambda_+(k(\dot{\gamma}^*\bar{s})) \lambda_-(k(\dot{\gamma}^*\bar{s})) \nu_1^*(k(\dot{\gamma}^*\bar{s}))k(\dot{\gamma}^*\bar{s})^2}{N_+(k(\dot{\gamma}^*\bar{s}))N_-(k(\dot{\gamma}^*\bar{s}))} \nonumber \\
&  &  +
2An_0^2 \sigma^{6} \int_0^{\infty} d\bar{t} 
e^{-\int_0^{\bar{t}} d\bar{s}  \lambda_+^*(\tilde{\bv{k}}(\dot{\gamma}^*\bar{s}))
 } 
e^{-\int_0^{\bar{t}} d\bar{s} \dot{\gamma}^*
\frac{2k_x\tilde{k}_y(\dot{\gamma}^*\bar{s})}
{\tilde{k}(\dot{\gamma}^*\bar{s})^2} \xi^{(1)} (k(\dot{\gamma}^*\bar{s}) )}
 \nonumber \\
  & & \times
\frac{k^2}{N_+(k)^2}
\frac{\lambda_+(k(\dot{\gamma}^*\bar{s}))^2  \nu_1^*(k(\dot{\gamma}^*\bar{s}))k(\dot{\gamma}^*\bar{s})^2}{N_+(k(\dot{\gamma}^*\bar{s}))^2} \nonumber \\
&  &  +
2An_0^2 \sigma^{6} \int_0^{\infty} d\bar{t} 
e^{-\int_0^{\bar{t}} d\bar{s}  \lambda_-^*(\tilde{\bv{k}}(\dot{\gamma}^*\bar{s}))
} 
e^{-\int_0^{\bar{t}} d\bar{s} \dot{\gamma}^*
\frac{2k_x\tilde{k}_y(\dot{\gamma}^*\bar{s})}
{\tilde{k}(\dot{\gamma}^*\bar{s})^2} \xi^{(2)} (k(\dot{\gamma}^*\bar{s}) )}
 \nonumber \\
  & & \times
\frac{k^2}{N_-(k)^2}
\frac{\lambda_-(k(\dot{\gamma}^*\bar{s}))^2  \nu_1^*(k(\dot{\gamma}^*\bar{s}))k(\dot{\gamma}^*\bar{s})^2}{N_-(k(\dot{\gamma}^*\bar{s}))^2}.
\label{C_nn:driv}
\end{eqnarray}

From the relations
\begin{eqnarray}
\frac{d}{d\bar{t}} e^{-\int_0^{\bar{t}} d\bar{s}  \nu_1^*(\tilde{\bv{k}}(\dot{\gamma}^*\bar{s}))
\tilde{k}(\dot{\gamma}^*\bar{s})^2 }
& = & -\nu_1^*(\tilde{\bv{k}}(\dot{\gamma}^*\bar{t}))
\tilde{k}(\dot{\gamma}^*\bar{t})^2
 e^{-\int_0^{\bar{t}} d\bar{s}  \nu_1^*(\tilde{\bv{k}}(\dot{\gamma}^*\bar{s}))
\tilde{k}(\dot{\gamma}^*\bar{s})^2 },
 \\
\frac{d}{d\bar{t}} 
e^{-\int_0^{\bar{t}} d\bar{s}  \lambda_+^*(\tilde{\bv{k}}(\dot{\gamma}^*\bar{s}))
}
& = & -
\lambda_+^*(\tilde{\bv{k}}(\dot{\gamma}^*\bar{t}))
e^{-\int_0^{\bar{t}} d\bar{s}  \lambda_+^*(\tilde{\bv{k}}(\dot{\gamma}^*\bar{s}))
},
 \\
\frac{d}{d\bar{t}} 
e^{-\int_0^{\bar{t}} d\bar{s}  \lambda_-^*(\tilde{\bv{k}}(\dot{\gamma}^*\bar{s}))
}
& = & -
\lambda_-^*(\tilde{\bv{k}}(\dot{\gamma}^*\bar{t}))
e^{-\int_0^{\bar{t}} d\bar{s}  \lambda_-^*(\tilde{\bv{k}}(\dot{\gamma}^*\bar{s}))},
\end{eqnarray}
eq. (\ref{C_nn:driv}) can be rewritten as
\begin{eqnarray}
\tilde{C}_{11}(\bv{k}) 
& = &  
An_0^2 \sigma^{6} 
\left \{
\frac{4k^2\lambda_+(k)\lambda_-(k)}{N_+(k)^2N_-(k)^2}
+\frac{k^4\lambda_+(k)\nu_1^*(k,e)}{N_+(k)^4}
+\frac{k^4\lambda_-(k)\nu_1^*(k,e)}{N_-(k)^4}
  \right \} \nonumber \\
  & & +
An_0^2 \sigma^{6} 
\int_0^{\infty} d\bar{t} 
\frac{4k^2\lambda_+(k)\lambda_-(k)}{N_+(k)N_-(k)}
e^{-\int_0^{\bar{t}} d\bar{s}  \nu_1^*(\tilde{\bv{k}}(\dot{\gamma}^*\bar{s}))
\tilde{k}(\dot{\gamma}^*\bar{s})^2 }  \nonumber \\
& & \times
\frac{d}{dt} \left \{
e^{-\int_0^{\bar{t}} d\bar{s} \dot{\gamma}^*
\frac{k_x\tilde{k}_y(\dot{\gamma}^*\bar{s})}
{\tilde{k}(\dot{\gamma}^*\bar{s})^2}
\{ 
  \xi^{(1)} (k(\dot{\gamma}^*\bar{s}) )
+ \xi^{(2)} (k(\dot{\gamma}^*\bar{s}) )
  \}} 
  G_{12}(k(\dot{\gamma}^*\bar{t}))
  \right \}
  \nonumber \\
  & & +
An_0^2 \sigma^{6} 
\int_0^{\infty} d\bar{t} 
\frac{k^4\lambda_+(k)\nu_1^*(k,e)}{N_+(k)^2}
e^{-\int_0^{\bar{t}} d\bar{s}  \lambda_+^*(\tilde{\bv{k}}(\dot{\gamma}^*\bar{s})) } 
 \nonumber \\
& & \times
\frac{d}{dt} \left \{
e^{-\int_0^{\bar{t}} d\bar{s} \dot{\gamma}^*
\frac{2k_x\tilde{k}_y(\dot{\gamma}^*\bar{s})}
{\tilde{k}(\dot{\gamma}^*\bar{s})^2}
  \xi^{(1)} (k(\dot{\gamma}^*\bar{s}) )
  } 
  G_{11}(k(\dot{\gamma}^*\bar{t}))
  \right \} \nonumber \\
  & & +
An_0^2 \sigma^{6} 
\int_0^{\infty} d\bar{t} 
\frac{k^4\lambda_-(k)\nu_1^*(k,e)}{N_-(k)^2}
e^{-\int_0^{\bar{t}} d\bar{s}  \lambda_-^*(\tilde{\bv{k}}(\dot{\gamma}^*\bar{s})) } 
 \nonumber \\
& & \times
\frac{d}{dt} \left \{
e^{-\int_0^{\bar{t}} d\bar{s} 
\frac{2k_x\tilde{k}_y(\dot{\gamma}^*\bar{s})}
{\tilde{k}(\dot{\gamma}^*\bar{s})^2}
  \xi^{(2)} (k(\dot{\gamma}^*\bar{s}) )
  } 
  G_{22}(k(\dot{\gamma}^*\bar{t}))
  \right \},
  \label{C11:ex}
\end{eqnarray}
where we have introduced
\begin{eqnarray}
G_{12}(k) & = & \frac{\lambda_+(k) \lambda_-(k)}{N_+(k) N_1(k)}, \\
G_{11}(k) & = & \frac{\lambda_+(k) \nu^*_1(k,e)k^2}{N_+(k)^2}, \\
G_{22}(k) & = & \frac{\lambda_-(k) \nu^*_1(k,e)k^2}{N_-(k)^2}.
\end{eqnarray}

It is possible to obtain the simpler expression for
$\tilde{C}_{11}(\bv{k})$ from the relation
$(4k^2\lambda_+\lambda_-/(N_+^2 N_-^2) + 
 \lambda_+ \nu_1^*(k,e)k^4/N_+^4 +\lambda_- \nu_1^*(k,e)k^4/N_-^4) = (\sigma^3 n_0 p^*(k,e))^{-1}
= S_0(k,e)/(\sigma^3 n_0 A)$,
and
\begin{eqnarray}
& & \frac{d}{dt} \left \{
e^{-\int_0^{\bar{t}} d\bar{s} \dot{\gamma}^*
\frac{k_x\tilde{k}_y(\dot{\gamma}^*\bar{s})}
{\tilde{k}(\dot{\gamma}^*\bar{s})^2}
\{ 
  \xi^{(1)} (k(\dot{\gamma}^*\bar{s}) )
+ \xi^{(2)} (k(\dot{\gamma}^*\bar{s}) )
  \}} 
  G_{12}(k(\dot{\gamma}^*\bar{t}))
  \right \}
  \nonumber \\
  &= & \dot{\gamma}^*
e^{-\int_0^{\bar{t}} d\bar{s} \dot{\gamma}^*
\frac{k_x\tilde{k}_y(\dot{\gamma}^*\bar{s})}
{\tilde{k}(\dot{\gamma}^*\bar{s})^2}
\{ 
  \xi^{(1)} (k(\dot{\gamma}^*\bar{s}) )
+ \xi^{(2)} (k(\dot{\gamma}^*\bar{s}) )
  \}} 
H_{12}(\tilde{k}(\dot{\gamma}^*\bar{t})),
\label{partial}
 \\
& & \frac{d}{dt} \left \{
e^{-\int_0^{\bar{t}} d\bar{s} \dot{\gamma}^*
\frac{2k_x\tilde{k}_y(\dot{\gamma}^*\bar{s})}
{\tilde{k}(\dot{\gamma}^*\bar{s})^2}
  \xi^{(1)} (k(\dot{\gamma}^*\bar{s}) )
  } 
  G_{11}(k(\dot{\gamma}^*\bar{t}))
  \right \}
  \nonumber \\
  &= & \dot{\gamma}^*
e^{-\int_0^{\bar{t}} d\bar{s} \dot{\gamma}^*
\frac{2k_x\tilde{k}_y(\dot{\gamma}^*\bar{s})}
{\tilde{k}(\dot{\gamma}^*\bar{s})^2}
  \xi^{(1)} (k(\dot{\gamma}^*\bar{s}) )
  } 
H_{11}(\tilde{k}(\dot{\gamma}^*\bar{t})),
 \\
& & \frac{d}{dt} \left \{
e^{-\int_0^{\bar{t}} d\bar{s} \dot{\gamma}^*
\frac{2k_x\tilde{k}_y(\dot{\gamma}^*\bar{s})}
{\tilde{k}(\dot{\gamma}^*\bar{s})^2}
  \xi^{(2)} (k(\dot{\gamma}^*\bar{s}) )
  } 
  G_{22}(k(\dot{\gamma}^*\bar{t}))
  \right \}
  \nonumber \\
  &= & \dot{\gamma}^*
e^{-\int_0^{\bar{t}} d\bar{s} 
\frac{2k_x\tilde{k}_y(\dot{\gamma}^*\bar{s})}
{\tilde{k}(\dot{\gamma}^*\bar{s})^2}
  \xi^{(2)} (k(\dot{\gamma}^*\bar{s}) )
  } 
H_{22}(\tilde{k}(\dot{\gamma}^*\bar{t})),
\end{eqnarray}
with
\begin{eqnarray}
H_{12}(k) & = & k \frac{dG_{12}(k)}{dk} - (\xi^{(1)}(k)+\xi^{(2)}(k))G_{12}(k),  \\
H_{11}(k) & = & k \frac{dG_{11}(k)}{dk} - 2 \xi^{(1)}(k) G_{11}(k),  \\
H_{22}(k) & = & k \frac{dG_{22}(k)}{dk} - 2 \xi^{(2)}(k) G_{22}(k).
\label{H:def}
\end{eqnarray}

Substituting (\ref{partial})-(\ref{H:def}) into eq. (\ref{C11:ex}),
we obtain
\begin{eqnarray}
\tilde{C}_{11}(\bv{k}) & = & n_0\sigma^3 \left \{
S_0(k,e) + \dot{\gamma}^* \tilde{\Delta}_1(\bv{k})
\right \}, \label{C11:fin}
 \end{eqnarray}
 where we have introduced
\begin{eqnarray}
\tilde{\Delta}_1(\bv{k}) & = &
 An_0 \sigma^3 \int_0^{\infty} d\bar{t} 
\frac{4k^2H_{12}(\tilde{k}(\dot{\gamma}^*\bar{t}))}{N_+(k) N_-(k)}
\frac{k_x\tilde{k}_y(\dot{\gamma}^*\bar{t})}{\tilde{k}(\dot{\gamma}{\bar{t}})^2}
e^{-\int_0^{\bar{t}} \bar{s} \{ \lambda^{(1)}(\tilde{\bv{k}}(\dot{\gamma}^*\bar{s}))
+ \lambda^{(2)}(\tilde{\bv{k}}(\dot{\gamma}^*\bar{s})) \}} \nonumber \\
& & +
An_0 \sigma^3  \int_0^{\infty} d\bar{t} \frac{k^2H_{11}(\tilde{k}(\dot{\gamma}^*\bar{t}))}{N_+(k)^2}\frac{k_x\tilde{k}_y(\dot{\gamma}^*\bar{t})}{\tilde{k}(\dot{\gamma}^*\bar{t})^2}e^{-2\int_0^{\bar{t}} d\bar{s} \lambda^{(1)}(\tilde{\bv{k}}(\dot{\gamma}^*\bar{s}))} \nonumber \\
& & +
An_0 \sigma^3  \int_0^{\infty} d\bar{t} \frac{k^2H_{22}(\tilde{k}(\dot{\gamma}^*\bar{t}))}{N_-(k)^2}\frac{k_x\tilde{k}_y(\dot{\gamma}^*\bar{t})}{\tilde{k}(\dot{\gamma}^*t)^2}e^{-2\int_0^{\bar{t}} d\bar{s} \lambda^{(2)}(\tilde{\bv{k}}(\dot{\gamma}^*\bar{s}))}.
 \end{eqnarray}
We, thus, obtain eq. (\ref{C_nn}) from eq. (\ref{C11:fin})
with
$\tilde{C}_{nn}(\bv{k},\bar{t}) = \sigma^{-3} \tilde{C}_{11}(\bv{k},\bar{t})$.

\section{Explicit representation of $\tilde{C}_{pp}(\bv{k})$}
\label{C_pp:app}

In this Appendix, 
we also explicitly derive eq. (\ref{C_pp}).
The momentum correlation
$ \tilde{C}_{pp}(\bv{k})$ can be written as
\begin{eqnarray}
 \tilde{C}_{pp}(\bv{k}) & = & (mn_0\sigma / t_E)^2 \{
 \tilde{C}_{22}(\bv{k}) + \tilde{C}_{33}(\bv{k}) +\tilde{C}_{44}(\bv{k})\},
 \label{C_pp:ex}
 \end{eqnarray}
 where
\begin{eqnarray}
 \tilde{C}_{22}(\bv{k}) + \tilde{C}_{33}(\bv{k}) +\tilde{C}_{44}(\bv{k})
 & = &
\int_0^{\infty} d\bar{t} \sum_{l,m} \sum_{\alpha=2,3,4}
  \tilde{\psi}^{(l)}_\alpha(\bv{k},\bar{t}) \tilde{\psi}^{(m)}_\alpha(-\bv{k},\bar{t})
F^{(lm)}(\tilde{\bv{k}}(\dot{\gamma}^*\bar{t})) \nonumber \\
 & = &
\int_0^{\infty} d\bar{t} \sum_{l,m}  \sum_{\alpha=2,3,4}
  (-1)^{\alpha+1} 
  \bar{\psi}^{(l)}_\alpha(\bv{k},\bar{t}) \bar{\psi}^{(m)}_\alpha(-\bv{k},\bar{t})
  \nonumber \\
  & & \times
F^{(lm)}(\tilde{\bv{k}}(\dot{\gamma}^*\bar{t})), 
 \label{C_uu:ex}
 \end{eqnarray}
 Here, to obtain (\ref{C_uu:ex})
we have used the relation $\tilde{\bv{\psi}}^{(l)}(\bv{k},\bar{t})
= ${\sf T}$ \cdot \bar{\bv{\psi}}^{(l)}(\bv{k},\bar{t})$ and 
$\sum_{\alpha=2,3,4} {\sf T}_{\alpha \beta}(\bv{k}) 
{\sf T}_{\alpha \gamma}(-\bv{k})
= (-1)^{\gamma} \sum_{\alpha=2,3,4} {\sf T}^{-1}_{\beta \alpha }(\bv{k}) 
{\sf T}_{\alpha \gamma}(\bv{k})
= (-1)^{\gamma+1} \delta_{\beta \gamma }$.

Introducing
 \begin{eqnarray}
 \tilde{D}_{\alpha }(\bv{k})
 & \equiv &
\int_0^{\infty} d\bar{t} \sum_{l,m} 
  \bar{\psi}^{(l)}_\alpha(\bv{k},\bar{t}) \bar{\psi}^{(m)}_\alpha(-\bv{k},\bar{t})
F^{(lm)}(\tilde{\bv{k}}(\dot{\gamma}^*\bar{t})), 
\label{D:def}
 \end{eqnarray}
 we can rewrite $\tilde{C}_{pp}(\bv{k})$ as
\begin{eqnarray}
 \tilde{C}_{pp}(\bv{k}) & = & (mn_0\sigma / t_E)^2 \{
 -\tilde{D}_{2}(\bv{k}) + \tilde{D}_{3}(\bv{k}) -\tilde{D}_{4}(\bv{k})\}.
 \end{eqnarray}

From (\ref{D:def}),
$\tilde{D}_{2}(\bv{k})$ can be rewritten as
\begin{eqnarray}
\tilde{D}_{2}(\bv{k}) & = &
4A \int_0^{\infty} d\bar{t} 
e^{-\int_0^{\bar{t}} d\bar{s}  \nu_1^*(\tilde{\bv{k}}(\dot{\gamma}^*\bar{s}))
\tilde{k}(\dot{\gamma}^*\bar{s})^2 } 
e^{-\int_0^{\bar{t}} d\bar{s} \dot{\gamma}^*
\frac{k_x\tilde{k}_y(\dot{\gamma}^*\bar{s})}
{\tilde{k}(\dot{\gamma}^*\bar{s})^2}
\{ 
  \xi^{(1)} (k(\dot{\gamma}^*\bar{s}) )
+ \xi^{(2)} (k(\dot{\gamma}^*\bar{s}) )
  \}} \nonumber \\
  & & \times
\frac{\lambda_+(k)\lambda_-(k)}{N_+(k)N_-(k)}
\frac{\lambda_+(k(\dot{\gamma}^*\bar{s})) \lambda_-(k(\dot{\gamma}^*\bar{s})) \nu_1^*(k(\dot{\gamma}^*\bar{s}))k(\dot{\gamma}^*\bar{s})^2}{N_+(k(\dot{\gamma}^*\bar{s}))N_-(k(\dot{\gamma}^*\bar{s}))} \nonumber \\
&  &  +
2A \int_0^{\infty} d\bar{t} 
e^{-\int_0^{\bar{t}} d\bar{s}  \lambda_+^*(\tilde{\bv{k}}(\dot{\gamma}^*\bar{s}))
 } 
e^{-\int_0^{\bar{t}} d\bar{s} \dot{\gamma}^*
\frac{2k_x\tilde{k}_y(\dot{\gamma}^*\bar{s})}
{\tilde{k}(\dot{\gamma}^*\bar{s})^2} \xi^{(1)} (k(\dot{\gamma}^*\bar{s}) )}
 \nonumber \\
  & & \times
\frac{\lambda_+(k)^2}{N_+(k)^2}
\frac{\lambda_+(k(\dot{\gamma}^*\bar{s}))^2  \nu_1^*(k(\dot{\gamma}^*\bar{s}))k(\dot{\gamma}^*\bar{s})^2}{N_+(k(\dot{\gamma}^*\bar{s}))^2} \nonumber \\
&  &  +
2A \int_0^{\infty} d\bar{t} 
e^{-\int_0^{\bar{t}} d\bar{s}  \lambda_-^*(\tilde{\bv{k}}(\dot{\gamma}^*\bar{s}))
} 
e^{-\int_0^{\bar{t}} d\bar{s} \dot{\gamma}^*
\frac{2k_x\tilde{k}_y(\dot{\gamma}^*\bar{s})}
{\tilde{k}(\dot{\gamma}^*\bar{s})^2} \xi^{(2)} (k(\dot{\gamma}^*\bar{s}) )}
 \nonumber \\
  & & \times
\frac{\lambda_-(k)^2}{N_-(k)^2}
\frac{\lambda_-(k(\dot{\gamma}^*\bar{s}))^2  \nu_1^*(k(\dot{\gamma}^*\bar{s}))k(\dot{\gamma}^*\bar{s})^2}{N_-(k(\dot{\gamma}^*\bar{s}))^2}.
\end{eqnarray}
From the parallel  procedure to that in Appendix \ref{C_nn:app},
we obtain
\begin{eqnarray}
\tilde{D}_{2}(\bv{k}) & = &
- A \left \{
\frac{4\lambda_+(k)^2\lambda_-(k)^2}{N_+(k)^2N_-(k)^2}
+ \frac{\lambda_+(k)^4}{N_+(k)^4}
+ \frac{\lambda_-(k)^4}{N_-(k)^4}
  \right \}
  - A \dot{\gamma}^*\tilde{\Delta}_2(\bv{k}),
\end{eqnarray}
where
\begin{eqnarray}
\tilde{\Delta}_2(\bv{k}) & = &
\int_0^{\infty} d\bar{t} 
\frac{4\lambda_+(k) \lambda_-(k)H_{12}(\tilde{k}(\dot{\gamma}^*\bar{t}))}{N_+(k) N_-(k)}\frac{k_x\tilde{k}_y(\dot{\gamma}^*\bar{t})}{\tilde{k}(\dot{\gamma}^*\bar{t})^2}e^{-\int_0^{\bar{t}} d\bar{s} \{ \lambda^{(1)}(\tilde{\bv{k}}(\dot{\gamma}^*\bar{s}))
+ \lambda^{(2)}(\tilde{\bv{k}}(\dot{\gamma}^*\bar{s})) \}} \nonumber \\
& & +
 \int_0^{\infty} d\bar{t} \frac{\lambda_+(k)^2 H_{11}(\tilde{k}(\dot{\gamma}^*\bar{t}))}{N_+(k)^2}\frac{k_x\tilde{k}_y(\dot{\gamma}^*\bar{t})}{\tilde{k}(\dot{\gamma}^*\bar{t})^2}e^{-2\int_0^{\bar{t}} d\bar{s} \lambda^{(1)}(\tilde{\bv{k}}(\dot{\gamma}^*\bar{s}))} \nonumber \\
& & +
 \int_0^{\infty} d\bar{t} \frac{\lambda_1(k)^2H_{22}(\tilde{k}(\dot{\gamma}^*\bar{t}))}
{N_-(k)^2}\frac{k_x\tilde{k}_y(\dot{\gamma}^*\bar{t})}{\tilde{k}(\dot{\gamma}^*\bar{t})^2}
e^{-2\int_0^{\bar{t}} d\bar{s} \lambda^{(2)}(\tilde{\bv{k}}(\dot{\gamma}^*\bar{s}))} .
 \end{eqnarray}
From $4\lambda_+(k)^2\lambda_-(k)^2/(N_+(k)^2N_-(k)^2)
+ \lambda_+(k)^4/N_+(k)^4
+ \lambda_-(k)^4/N_-(k)^4 = 1$,
we find that $\tilde{D}_{2}(\bv{k})$ is represented by eq. (\ref{D:2}).

From (\ref{D:def}),
we obtain
\begin{eqnarray}
\tilde{D}_{3}(\bv{k}) & = &
2A \int_0^{\infty} d\bar{t} 
e^{-\int_0^{\bar{t}} d\bar{s}  2\nu_2^*(\tilde{\bv{k}}(\dot{\gamma}^*\bar{s}))
\tilde{k}(\dot{\gamma}^*\bar{s})^2 } 
e^{\int_0^{\bar{t}} d\bar{s} \dot{\gamma}^*
2\frac{k_x\tilde{k}_y(\dot{\gamma}^*\bar{s})}
{\tilde{k}(\dot{\gamma}^*\bar{s})^2} } 
 \nu_2^*(k(\dot{\gamma}^*\bar{s}))k(\dot{\gamma}^*\bar{s})^2.
 \end{eqnarray}
 From the identities
\begin{eqnarray}
\frac{d}{d\bar{t}} e^{-\int_0^{\bar{t}} d\bar{s}  \nu_2^*(\tilde{\bv{k}}(\dot{\gamma}^*\bar{s}))
\tilde{k}(\dot{\gamma}^*\bar{s})^2 }
& = & -\nu_2^*(\tilde{\bv{k}}(\dot{\gamma}^*\bar{t}))
\tilde{k}(\dot{\gamma}^*\bar{t})^2
 e^{-\int_0^{\bar{t}} d\bar{s}  \nu_2^*(\tilde{\bv{k}}(\dot{\gamma}^*\bar{s}))
\tilde{k}(\dot{\gamma}^*\bar{s})^2 },
\label{exp:nu2}
\end{eqnarray}
and
\begin{eqnarray}
e^{\int_0^{\bar{t}} d\bar{s} \dot{\gamma}^*
\frac{k_x\tilde{k}_y(\dot{\gamma}^*\bar{s})}
{\tilde{k}(\dot{\gamma}^*\bar{s})^2} }
= \frac{\tilde{k}(\dot{\gamma}^*\bar{s})}{k^2},
\end{eqnarray}
we find that $\tilde{D}_{3}(\bv{k})$ is expressed as eq. (\ref{D:3})
with
\begin{eqnarray}
 \tilde{\Delta}_3(\bv{k}) & = &
\int_0^{\infty} d\bar{t} 
\frac{k_x \tilde{k}_y(\dot{\gamma}^*\bar{t})}{k^2}e^{-\int_0^{\bar{t}} d\bar{s} 2\nu^*_2(\tilde{k}(\dot{\gamma}^*\bar{s}))\tilde{k}(\dot{\gamma}^*\bar{s})^2}.
\end{eqnarray}

From eq. (\ref{D:def}),
we obtain
\begin{eqnarray}
\tilde{D}_{4}(\bv{k}) & = &
-2A \int_0^{\infty} d\bar{t} 
\left \{
M^2(\bv{k}) \frac{\tilde{k}(\dot{\gamma}^*\bar{s})^2}{k^2}
-2M(\bv{k})M(\tilde{\bv{k}}(\dot{\gamma}^*\bar{s}))
\frac{\tilde{k}(\dot{\gamma}^*\bar{s})}{k}
+M(\tilde{\bv{k}}(\dot{\gamma}^*\bar{s}))^2 + 1
  \right \}
  \nonumber \\
  & & \times
e^{-\int_0^{\bar{t}} d\bar{s}  2\nu_2^*(\tilde{\bv{k}}(\dot{\gamma}^*\bar{s}))
\tilde{k}(\dot{\gamma}^*\bar{s})^2 } 
 \nu_2^*(k(\dot{\gamma}^*\bar{t}))k(\dot{\gamma}^*\bar{t})^2.
 \end{eqnarray}
With the aid of (\ref{exp:nu2}),
we confirm that $\tilde{D}_{4}(\bv{k})$ can be represented by eq. (\ref{D:4}),
where
\begin{eqnarray}
 \tilde{\Delta}_4(\bv{k}) & = &
\int_0^{\infty} d\bar{t} \left \{ 
\frac{k_x \tilde{k}_y(\dot{\gamma}^*\bar{t})}{\tilde{k}(\dot{\gamma}^*\bar{t})^2}F(\bv{k},\bar{t})- \frac{k_z}{\tilde{k}(\dot{\gamma}^*\bar{t})} \right \}F(\bv{k},\bar{t})
e^{-\int_0^{\bar{t}} d\bar{s} 2\nu^*_2(\tilde{k}(\dot{\gamma}^*\bar{s}))\tilde{k}(\dot{\gamma}^*\bar{s})^2},
\end{eqnarray}
and
$F(\bv{k},t) = M(\hat{\bv{k}}(t)) - \hat{k}(t) M(k)/k$.

Finally, substituting eqs. (\ref{D:2}), (\ref{D:3}) and (\ref{D:4}) into (\ref{C_pp:ex}) with 
(\ref{C_uu:ex}), we obtain the 
explicit representation of $\tilde{C}_{pp}(\bv{k})$ as eq. (\ref{C_pp}).

\section{Evaluation of Asymptotic Form of $\bar{\Delta}_j(r)$}

In this Appendix, we evaluate the asymptotic form of $\bar{\Delta}_j(r)$
for $r \gg l_c$.

\subsection{The Asymptotic Form of $\bar{\Delta}_1(r)$}

We note that  $\bar{\Delta}_1(r)$ can be separated into three parts 
 \begin{eqnarray}
 \bar{\Delta}_1(r) & = & \bar{\Delta}_{11}(r) +  \bar{\Delta}_{12}(r) +\bar{\Delta}_{13}(r),  
\end{eqnarray}
where
\begin{eqnarray}
\bar{\Delta}_{11}(r) & = &
 \frac{A}{n_0 \sigma^6} \int \frac{d \bv{k}}{(2\pi)^3}
 \frac{\sin{(kr/\sigma)}}{kr/\sigma} 
 \int_0^{\infty} d\bar{t} 
\frac{4k^2H_{12}(\tilde{k}(\dot{\gamma}^*\bar{t}))}{N_+(k) N_-(k)}
\frac{k_x\tilde{k}_y(\dot{\gamma}^*\bar{t})}{\tilde{k}(\dot{\gamma}^*{\bar{t}})^2}
\nonumber \\
& & \times
e^{-\int_0^{\bar{t}} \bar{s} \{ \lambda^{(1)}(\tilde{\bv{k}}(\dot{\gamma}^*\bar{s}))
+ \lambda^{(2)}(\tilde{\bv{k}}(\dot{\gamma}^*\bar{s})) \}}, \label{D11}\\
\bar{\Delta}_{12}(r) & = &
 \frac{A}{n_0 \sigma^6} \int \frac{d \bv{k}}{(2\pi)^3}
 \frac{\sin{(kr/\sigma)}}{kr/\sigma} 
 \int_0^{\infty} d\bar{t} \frac{k^2H_{11}(\tilde{k}(\dot{\gamma}^*\bar{t}))}{N_+(k)^2}\frac{k_x\tilde{k}_y(\dot{\gamma}^*\bar{t})}{\tilde{k}(\dot{\gamma}^*\bar{t})^2}e^{-2\int_0^{\bar{t}} d\bar{s} \lambda^{(1)}(\tilde{\bv{k}}(\dot{\gamma}^*\bar{s}))}, \label{D12}\\
\bar{\Delta}_{13}(r) & = &
 \frac{A}{n_0 \sigma^6} \int \frac{d \bv{k}}{(2\pi)^3}
 \frac{\sin{(kr/\sigma)}}{kr/\sigma} 
\int_0^{\infty} d\bar{t} \frac{k^2H_{22}(\tilde{k}(\dot{\gamma}^*\bar{t}))}{N_-(k)^2}\frac{k_x\tilde{k}_y(\dot{\gamma}^*\bar{t})}{\tilde{k}(\dot{\gamma}^*t)^2}e^{-2\int_0^{\bar{t}} d\bar{s} \lambda^{(2)}(\tilde{\bv{k}}(\dot{\gamma}^*\bar{s}))}. 
\label{D13}
\end{eqnarray}
Let us introduce $l_c \equiv \sigma / \sqrt{\dot{\gamma}^*}$,
$r^* \equiv r / l_c$, and the transformation
 $\bar{t} = \tau / \dot{\gamma}^*$,  
$\bar{s} = \tau' / \dot{\gamma}^*$, 
$\bv{k} = \sqrt{\dot{\gamma}^*} \bv{K} / r^* = \sigma \bv{K} /r$, and
$\tilde{\bv{k}}(\tau) = \sqrt{\dot{\gamma}^*} \tilde{\bv{K}}(\tau) / r^*$.
By using the new variables, 
eqs. (\ref{D11}), (\ref{D12}), and (\ref{D13}) can be rewritten as
\begin{eqnarray}
\bar{\Delta}_{11}(r) & = & r^{*-5}
 \frac{A}{n_0 \sigma^3 l_c^{3} } \int \frac{d \bv{K}}{(2\pi)^3}
 \frac{\sin{K}}{K} 
 \int_0^{\infty} d\tau 
\frac{4K^2H_{12}(\sqrt{\dot{\gamma}^*}\tilde{K}(\tau)/ r^*)}{N_+(\sqrt{\dot{\gamma}^*}K/ r^*) N_-(\sqrt{\dot{\gamma}^*}K/ r^*)}
\frac{K_x\tilde{K}_y(\tau)}{\tilde{K}(\tau)^2} \nonumber \\
& & \times
e^{-\int_0^{\tau} \tau' \dot{\gamma}^{*-1} \{ \lambda^{(1)}(\sqrt{\dot{\gamma}^*}\tilde{\bv{K}}(\tau')/ r^*)
+ \lambda^{(2)}(\sqrt{\dot{\gamma}^*}\tilde{\bv{K}}(\tau')/ r^*) \}}, \label{D11:ex} \\
\bar{\Delta}_{12}(r) & = &r^{*-5}
 \frac{A}{n_0 \sigma^3 l_c^{3} } \int \frac{d \bv{K}}{(2\pi)^3}
 \frac{\sin{K}}{K} 
 \int_0^{\infty} d\tau \frac{K^2H_{11}(\sqrt{\dot{\gamma}^*}\tilde{K}(\tau)/ r^*)}{N_+(\sqrt{\dot{\gamma}^*}K/ r^*)^2}\frac{K_x\tilde{K}_y(\tau)}{\tilde{K}(\tau)^2}\nonumber \\
 & & \times e^{-2\int_0^{\tau} d\tau' \dot{\gamma}^{*-1}\lambda^{(1)}(\sqrt{\dot{\gamma}^*}\tilde{\bv{K}}(\tau')/ r^*)}, \label{D12:ex} \\
\bar{\Delta}_{13}(r) & = &r^{*-5}
 \frac{A}{n_0 \sigma^3 l_c^{3} } \int \frac{d \bv{K}}{(2\pi)^3}
 \frac{\sin{K}}{K} 
\int_0^{\infty} d\tau \frac{K^2H_{22}(\sqrt{\dot{\gamma}^*}\tilde{K}(\tau)/ r^*)}{N_-(\sqrt{\dot{\gamma}^*}K/ r^*)^2}\frac{K_x\tilde{K}_y(\tau)}{\tilde{K}(\tau)^2} \nonumber \\
& & \times e^{-2\int_0^{\tau} d\tau' \dot{\gamma}^{*-1} \lambda^{(2)}(\sqrt{\dot{\gamma}^*}\tilde{\bv{k}}(\tau')/ r^*)}. \label{D13:ex}
\end{eqnarray}

Introducing the transformations $\tau = r^{*2/3} Y$ and $\tau' = r^{*2/3} Y'$
to extract the asymptotic forms for $r^* \gg 1$,
$\bar{\Delta}_{1j}(r)$  ($j=1,2,3$) can be rewritten as
\begin{eqnarray}
\bar{\Delta}_{11}(r) & = & r^{*-13/3}
 \frac{A}{n_0 \sigma^3 l_c^{3} } \int \frac{d \bv{K}}{(2\pi)^3}
 \frac{\sin{K}}{K} 
 \int_0^{\infty} dY 
\frac{4K^2H_{12}(\sqrt{\dot{\gamma}^*}\tilde{K}(r^{*2/3} Y)/ r^*)}{N_+(\sqrt{\dot{\gamma}^*}K/ r^*) N_-(\sqrt{\dot{\gamma}^*}K/ r^*)}
\frac{K_x\tilde{K}_y(r^{*2/3} Y)}{\tilde{K}(r^{*2/3} Y)^2} \nonumber \\
& & \times
e^{-\int_0^{\tau} dY' \dot{\gamma}^{*-1} r^{*2/3}\{ \lambda^{(1)}(\sqrt{\dot{\gamma}^*}\tilde{\bv{K}}(r^{*2/3} Y')/ r^*)
+ \lambda^{(2)}(\sqrt{\dot{\gamma}^*}\tilde{\bv{K}}(r^{*2/3} Y')/ r^*) \}}, \label{Delta11:e} \\
\bar{\Delta}_{12}(r) & = &r^{*-13/3}
 \frac{A}{n_0 \sigma^3 l_c^{3} } \int \frac{d \bv{K}}{(2\pi)^3}
 \frac{\sin{K}}{K} 
 \int_0^{\infty} dY \frac{K^2H_{11}(\sqrt{\dot{\gamma}^*}\tilde{K}(r^{*2/3} Y)/ r^*)}{N_+(\sqrt{\dot{\gamma}^*}K/ r^*)^2}\frac{K_x\tilde{K}_y(r^{*2/3} Y)}{\tilde{K}(r^{*2/3} y)^2}\nonumber \\
 & & \times e^{-2\int_0^{\tau} dY' \dot{\gamma}^{*-1} r^{*2/3} \lambda^{(1)}(\sqrt{\dot{\gamma}^*}\tilde{\bv{K}}(r^{*2/3} Y')/ r^*)}, \label{Delta12:e} \\
\bar{\Delta}_{13}(r) & = &r^{*-13/3}
 \frac{A}{n_0 \sigma^3 l_c^{3} } \int \frac{d \bv{K}}{(2\pi)^3}
 \frac{\sin{K}}{K} 
\int_0^{\infty} dY \frac{K^2H_{22}(\sqrt{\dot{\gamma}^*}\tilde{K}(r^{*2/3} Y)/ r^*)}{N_-(\sqrt{\dot{\gamma}^*}K/ r^*)^2}\frac{K_x\tilde{K}_y(r^{*2/3} Y)}{\tilde{K}(r^{*2/3} Y)^2} \nonumber \\
& & \times e^{-2\int_0^{\tau} dY' \dot{\gamma}^{*-1} r^{*2/3} \lambda^{(2)}(\sqrt{\dot{\gamma}^*}\tilde{\bv{k}}(r^{*2/3} Y')/ r^*)}, \label{Delta13:e}
\end{eqnarray}
where the functions included in (\ref{Delta11:e}), (\ref{Delta12:e}), and 
(\ref{Delta13:e}) have the asymptotic forms for $r^* \gg 1$ 
 \begin{eqnarray}
\tilde{K}_y(\bar{r}^{2/3} Y) & \simeq &  r^{*2/3}  \{ YK_x + O(r^{*-2/3}) \},
 \label{k:asym1}\\
\tilde{K}(r^{*2/3} Y')^2
& \simeq & r^{*4/3}\{ Y^{'2} K_x^2 + O(r^{*-2/3})\},
\label{k:asym2}
\end{eqnarray}
 \begin{eqnarray}
H_{12}(\sqrt{\dot{\gamma}^*}\tilde{K}(r^{*2/3} Y)/ r^*) & \simeq &
\frac{1}{2} + O(r^{*-2/3}), \label{H:asym1} \\
H_{11}(\sqrt{\dot{\gamma}^*}\tilde{K}(r^{*2/3} Y)/ r^*) & \simeq &
\frac{i \sqrt{\dot{\gamma}^*} \nu^*_1(0,t)}{2c_0}  Y |K_x| r^{*-1/3} + O(r^{*-2/3}), \label{H:asym2} \\
H_{22}(\sqrt{\dot{\gamma}^*}\tilde{K}(r^{*2/3} Y)/ r^*) & \simeq &
-\frac{i \sqrt{\dot{\gamma}^*} \nu^*_1(0,t)}{2c_0}  y |K_x| r^{*-1/3} + O(r^{*-2/3}), 
\label{H:asym3}
\end{eqnarray}
 \begin{eqnarray}
N_+(\sqrt{\dot{\gamma}^*}K/ r^*) & \simeq &
\sqrt{2\dot{\gamma}^*}K i c_0 r^{*-1}
 + O(r^{*-2}),  \\
 N_-(\sqrt{\dot{\gamma}^*}K/ r^*) & \simeq &
\sqrt{2\dot{\gamma}^*}K i c_0 r^{*-1}
 + O(r^{*-2}), 
 \label{N:asym}
  \end{eqnarray}
 \begin{eqnarray}
e^{-\int_0^{\tau} dY' \dot{\gamma}^{*-1} r^{*2/3} \lambda^{(1)}(\sqrt{\dot{\gamma}^*}\tilde{\bv{K}}(r^{*2/3} Y')/ r^*)} & \simeq &
e^{-i c_0 \dot{\gamma}^{*-1/2} \int_0^{Y} dy' |K_x| Y' r^{*1/2}} 
\sqrt{\frac{K}{|K_x|r^{*2/3}Y}} \nonumber \\
& & \times e^{-\int_0^{y} dY' \nu^*_1(0,t) K_x^2 Y{'2}}
\left \{
   1
+ O(r^{*-2/3})  
\right \},  \\
e^{-\int_0^{\tau} dY' \dot{\gamma}^{*-1} r^{*2/3} \lambda^{(2)}(\sqrt{\dot{\gamma}^*}\tilde{\bv{K}}(r^{*2/3} Y')/ r^*)} & \simeq &
e^{i c_0 \dot{\gamma}^{*-1/2} \int_0^{Y} dY' |K_x| Y' r^{*1/2}} 
\sqrt{\frac{K}{|K_x|r^{*2/3}Y}} \nonumber \\
& & \times e^{-\int_0^{Y} dY' \nu^*_1(0,t) K_x^2 Y{'2}}
\left \{
   1
+ O(r^{*-2/3})  
\right \}, 
\label{e:asym}
  \end{eqnarray}
with $c_0^2 \equiv n_0\sigma^3 p^*(0,e)$.
Substituting these relations into eqs. (\ref{Delta11:e})-(\ref{Delta13:e}),
we obtain
\begin{eqnarray}
\bar{\Delta}_{11}(r) & = & J_{11} r^{*-11/3}
 + O(r^{*-13/3}),
\nonumber \\
|\bar{\Delta}_{12}(r) | & = &|\bar{\Delta}_{13}(r) | = J_{12}r^{*-4}
 + O(r^{*-14/3}),
\end{eqnarray}
where we have introduced
\begin{eqnarray}
J_{11} & = & 
 \frac{A}{\dot{\gamma}^*c_0 n_0 \sigma^3 l_c^{3} } \int \frac{d \bv{K}}{(2\pi)^3}
 \frac{\sin{K}}{K} 
 \int_0^{\infty} dY
\frac{1}{Y^2}
\frac{K}{|K_x|} 
e^{-2\nu_1^*(0) Y^{3}/3},
 \\
J_{12}  & = &
 \frac{A}{4n_0 \sigma^3 l_c^{3} \sqrt{\dot{\gamma}^*}c_0^3 } \int \frac{d \bv{K}}{(2\pi)^3}
 \frac{\sin{K}}{K} 
 \int_0^{\infty} dY \frac{K}{Y} e^{-2 \nu_1^*(0) Y^{3}/3}. 
\end{eqnarray}
Thus, the asymptotic form of $\bar{\Delta}_1(r)$ can be represented by
eq. (\ref{Delta1:asym}).

\subsection{The Asymptotic Form of $\bar{\Delta}_2(r)$}

Similarly, $\bar{\Delta}_2(r)$ is given by summation of the three terms
 \begin{eqnarray}
 \bar{\Delta}_2(r) & = & \bar{\Delta}_{21}(r) +  \bar{\Delta}_{22}(r) +\bar{\Delta}_{23}(r),  
\end{eqnarray}
where
\begin{eqnarray}
\bar{\Delta}_{21}(r) & = & \sigma^{-3} \int \frac{d \bv{k}}{(2\pi)^3}
 \frac{\sin{(kr/\sigma)}}{kr/\sigma} 
 \int_0^{\infty} d\bar{t} 
\frac{4\lambda_+(k) \lambda_-(k)H_{12}(\tilde{k}(\dot{\gamma}^*\bar{t}))}{N_+(k) N_-(k)}\frac{k_x\tilde{k}_y(\dot{\gamma}^*\bar{t})}{\tilde{k}(\dot{\gamma}^*\bar{t})^2} \nonumber \\
& & \times e^{-\int_0^{\bar{t}} d\bar{s} \{ \lambda^{(1)}(\tilde{\bv{k}}(\dot{\gamma}^*\bar{s}))
+ \lambda^{(2)}(\tilde{\bv{k}}(\dot{\gamma}^*\bar{s})) \}},
  \\
\bar{\Delta}_{22}(r) & = &\sigma^{-3} \int \frac{d \bv{k}}{(2\pi)^3}
 \frac{\sin{(kr/\sigma)}}{kr/\sigma} 
 \int_0^{\infty} d\bar{t} \frac{\lambda_+(k)^2 H_{11}(\tilde{k}(\dot{\gamma}^*\bar{t}))}{N_+(k)^2}\frac{k_x\tilde{k}_y(\dot{\gamma}^*\bar{t})}{\tilde{k}(\dot{\gamma}^*\bar{t})^2} \nonumber \\
 & & \times e^{-2\int_0^{\bar{t}} d\bar{s} \lambda^{(1)}(\tilde{\bv{k}}(\dot{\gamma}^*\bar{s}))} ,
   \\
\bar{\Delta}_{23}(r) & = &\sigma^{-3} \int \frac{d \bv{k}}{(2\pi)^3}
 \frac{\sin{(kr/\sigma)}}{kr/\sigma} 
 \int_0^{\infty} d\bar{t} \frac{\lambda_1(k)^2H_{22}(\tilde{k}(\dot{\gamma}^*\bar{t}))}
{N_-(k)^2}\frac{k_x\tilde{k}_y(\dot{\gamma}^*\bar{t})}{\tilde{k}(\dot{\gamma}^*\bar{t})^2}
\nonumber \\
& & \times 
e^{-2\int_0^{\bar{t}} d\bar{s} \lambda^{(2)}(\tilde{\bv{k}}(\dot{\gamma}^*\bar{s}))}. 
\end{eqnarray}
As in the case of the previous subsection,
from the transformations $\bar{t} =r^{*2/3} Y / \dot{\gamma}^*$,  
$\bar{s} = r^{*2/3} Y' / \dot{\gamma}^*$, 
and $\bv{k} = \sqrt{\dot{\gamma}^*} \bv{K} / r^*$,
we obtain
\begin{eqnarray}
\bar{\Delta}_{21}(r) & = &  \frac{r^{*-7/3}}{\dot{\gamma}^{*} l_c^3} \int \frac{d \bv{K}}{(2\pi)^3}
 \frac{\sin{K}}{K} 
 \int_0^{\infty} dY
\frac{4\lambda_+(\sqrt{\dot{\gamma}^*} K / r^*) \lambda_-(\sqrt{\dot{\gamma}^*} K / r^*)H_{12}(\sqrt{\dot{\gamma}^*}\tilde{K}(r^{*2/3} Y)/ r^*)}{N_+(\sqrt{\dot{\gamma}^*} K / r^*) N_-(\sqrt{\dot{\gamma}^*} K / r^*)} \nonumber \\
& & \times \frac{K_x\tilde{K}_y(r^{*2/3} Y)}{\tilde{K}(r^{*2/3} Y)^2} e^{-\int_0^{Y} dY' \dot{\gamma}^{*-1}r^{*2/3}\{ \lambda^{(1)}(\sqrt{\dot{\gamma}^*}\tilde{\bv{K}}(r^{*2/3} Y')/ r^*)
+ \lambda^{(2)}(\sqrt{\dot{\gamma}^*}\tilde{\bv{K}}(r^{*2/3} Y')/ r^*) \}},
\label{Delta_21}
  \\
\bar{\Delta}_{22}(r) & = & \frac{r^{*-7/3}}{\dot{\gamma}^{*} l_c^3}  \int \frac{d \bv{K}}{(2\pi)^3}
 \frac{\sin{K}}{K} 
 \int_0^{\infty} dY \frac{\lambda_+(\sqrt{\dot{\gamma}^*} K / r^*)^2 H_{11}(\sqrt{\dot{\gamma}^*} \tilde{K}(r^{*2/3} Y) / r^*)}{N_+(\sqrt{\dot{\gamma}^*} \bv{K} / r^*)^2}\frac{K_x\tilde{K}_y((r^{*2/3} Y)}{\tilde{K}((r^{*2/3} Y)^2} \nonumber \\
 & & \times e^{-2\int_0^{Y} dY' \dot{\gamma}^{*-1}r^{*2/3}\lambda^{(1)}(\sqrt{\dot{\gamma}^*} \tilde{\bv{K}}(r^{*2/3} Y') / r^*)}, 
\label{Delta_22}
   \\
\bar{\Delta}_{23}(r) & = & \frac{r^{*-7/3}}{\dot{\gamma}^{*} l_c^3} \int \frac{d \bv{K}}{(2\pi)^3}
 \frac{\sin{K}}{K} 
 \int_0^{\infty} dY \frac{\lambda_-(\sqrt{\dot{\gamma}^*} K / r^*)^2H_{22}(\sqrt{\dot{\gamma}^*} \tilde{K}(r^{*2/3} Y) / r^*)}
{N_-(\sqrt{\dot{\gamma}^*} K / r^*)^2}\frac{K_x\tilde{K}_y(r^{*2/3} Y)}{\tilde{K}(r^{*2/3} Y)^2}
\nonumber \\
& & \times
e^{-2\int_0^{Y} dY' \dot{\gamma}^{*-1}\lambda^{(2)}r^{*2/3}(\sqrt{\dot{\gamma}^*} \tilde{\bv{K}}(r^{*2/3} Y') / r^*)}. 
\label{Delta_23}
\end{eqnarray}

From eqs. (\ref{k:asym1})-(\ref{e:asym}) and eqs. 
(\ref{Delta_21})-(\ref{Delta_23}),
we obtain
\begin{eqnarray}
\bar{\Delta}_{21}(r) & = & \frac{c_0 n_0 \sigma^3}{A}J_{11} r^{*-11/3}
 + O(r^{*-13/3}),
 \\
|\bar{\Delta}_{22}(r) | & = &|\bar{\Delta}_{23}(r) | = \frac{c_0 n_0 \sigma^3}{A}J_{12}r^{*-4}
 + O(r^{*-14/3}). 
\end{eqnarray}
We, thus, find the asymptotic form of $\bar{\Delta}_2(r)$ is given by
eq. (\ref{Delta2:asym}).

\subsection{Asymptotic Form of $\bar{\Delta}_3(r)$}

 $\bar{\Delta}_3(r)$ can be represented by
\begin{eqnarray}
 \bar{\Delta}_3(r) & = & \sigma^{-3} \int \frac{d \bv{k}}{(2\pi)^3}
 \frac{\sin{(kr/\sigma)}}{kr/\sigma} 
\int_0^{\infty} d\bar{t} 
\frac{k_x \tilde{k}_y(\dot{\gamma}^*\bar{t})}{k^2}e^{-\int_0^{\bar{t}} d\bar{s} 2\nu^*_2(\tilde{k}(\dot{\gamma}^*\bar{s}),0)\tilde{k}(\dot{\gamma}^*\bar{s})^2}.  
\label{Delta3:def}
\end{eqnarray}
From the transformations
$\bar{t} = \tau / \dot{\gamma}^*$,  
$\bar{s} = \tau' / \dot{\gamma}^*$, 
and $\bv{k} = \sqrt{\dot{\gamma}^*} \bv{K} / \bar{r}$,
we can rewritten (\ref{Delta3:def}) as
\begin{eqnarray}
 \bar{\Delta}_3(r) & = & \dot{\gamma}^{*-1} l_c^{-3} r^{*-3}
 \int \frac{d \bv{K}}{(2\pi)^3}
 \frac{\sin{K}}{K} 
\int_0^{\infty} d\tau 
\frac{K_x \tilde{K}_y(\tau)}{K^2}e^{-\int_0^{\tau} d\tau' 2\nu^*_2 
\left ( \frac{\sqrt{\dot{\gamma}^*}\tilde{K}(\tau)}{ r^*} \right )
\frac{\tilde{K}(\tau)^2}{ r^{*2}}}.  
\end{eqnarray}
Introducing the other transformations
$\tau = r^{*2/3} y$ and $\tau' = r^{*2/3} y'$,
we obtain
 \begin{eqnarray}
 \bar{\Delta}_3(r) & = & \dot{\gamma}^{*-1} l_c^{-3} r^{*-7/3}
 \int \frac{d \bv{K}}{(2\pi)^3}
 \frac{\sin{K}}{K} 
\int_0^{\infty} dy 
\frac{K_x \tilde{K}_y(r^{*2/3} y)}{K^2} \nonumber \\
& & \times e^{-\int_0^{y} dy' 2\nu^*_2 
\left ( \frac{\sqrt{\dot{\gamma}^*}\tilde{K}(r^{*2/3} y')}{ r^*} \right )
\frac{\tilde{K}(r^{*2/3} y')^2}{ r^{*4/3}}}.  \label{Delta:trans}
\end{eqnarray}

We should note that 
\begin{eqnarray}
\nu^*_2 
\left ( \frac{\sqrt{\dot{\gamma}^*}\tilde{K}(r^{*2/3} y')}{ r^*} \right )
\simeq \nu^*_2(0) + O(r^{*-4/3})
\end{eqnarray}
can be used for $r^*\gg 1$ from eq. (\ref{k:asym1}).
Using this equation and eq. (\ref{k:asym2}),
we obtain
 \begin{eqnarray}
 \bar{\Delta}_3(r) & = & J_3   r^{*-5/3}
+ O(r^{*-2/3}),
\end{eqnarray}
where
 \begin{eqnarray}
 J_3 & = & \dot{\gamma}^{*-1} l_c^{-3}
 \int \frac{d \bv{K}}{(2\pi)^3}
 \frac{\sin{K}}{K} 
\int_0^{\infty} dY 
\frac{K_x^2 Y}{K^2}e^{-  2\nu^*_2(0)K_x^2 Y^{3}/3}.
\end{eqnarray}
Therefore, the asymptotic form of $\bar{\Delta}_3(r)$ is given
by eq. (\ref{Delta3:asym}).

\subsection{The Asymptotic Form of $\bar{\Delta}_4(r)$}

 $\bar{\Delta}_4(r)$ can be rewritten as
\begin{eqnarray}
 \tilde{\Delta}_4(\bv{k}) & = &\sigma^{-3}\int \frac{d \bv{k}}{(2\pi)^3}
 \frac{\sin{(kr/\sigma)}}{kr/\sigma} 
\int_0^{\infty} d\bar{t} F_4(\bv{k},\bar{t})
e^{-\int_0^{\bar{t}} d\bar{s} 2\nu^*_2(\tilde{k}(\dot{\gamma}^*\bar{s}))\tilde{k}(\dot{\gamma}^*\bar{s})^2},
\end{eqnarray}
where we have introduced
$F_4(\bv{k},t) = \left \{ 
\frac{k_x \hat{k}_y(\dot{\gamma}^*\bar{t})}{\hat{k}(\dot{\gamma}^*\bar{t})^2}F(\bv{k},\bar{t})- \frac{k_z}{\hat{k}(\dot{\gamma}^*\bar{t})} \right \}F(\bv{k},\bar{t}).$
From the transformations $\bar{t} =r^{*2/3} Y / \dot{\gamma}^*$,  
$\bar{s} = r^{*2/3} Y' / \dot{\gamma}^*$, 
and $\bv{k} = \sqrt{\dot{\gamma}^*} \bv{K} / r^*$,
we obtain
\begin{eqnarray}
 \tilde{\Delta}_4(\bv{k}) & = &\dot{\gamma}^{*-1} l_c^{-3} r^{*-7/3}
 \int \frac{d \bv{K}}{(2\pi)^3}
 \frac{\sin{K}}{K} 
\int_0^{\infty} dY 
F_4(\sqrt{\dot{\gamma}^*} \bv{K} / r^*,r^{*2/3} Y / \dot{\gamma}^*) \nonumber \\
& & \times e^{-\int_0^{Y} dY' 2\nu^*_2 
\left ( \frac{\sqrt{\dot{\gamma}^*}\tilde{K}(r^{*2/3} Y')}{ r^*} \right )
\frac{\tilde{K}(r^{*2/3} Y')^2}{ r^{*4/3}}}. 
\label{Delta4:e}
\end{eqnarray}

Substituting the asymptotic expression
 \begin{eqnarray}
F_4(\sqrt{\dot{\gamma}^*} \bv{K} / r^*,r^{*2/3} Y / \dot{\gamma}^*)
\simeq \left ( 
\frac{\pi K_Z}{2K_\perp} + \frac{|K_x|}{K} M(\bv{K}) 
\right )^2 r^{*2/3}Y
  \end{eqnarray}
for $r^*\gg 1$,
eqs. (\ref{k:asym1}) and (\ref{k:asym2})
into eq. (\ref{Delta4:e}),
we obtain
 \begin{eqnarray}
 \bar{\Delta}_4(r) & = & J_4   r^{*-5/3}
+ O(r^{*-2/3}),
\end{eqnarray}
where
 \begin{eqnarray}
 J_4 & = & \dot{\gamma}^{*-1} l_c^{-3}
 \int \frac{d \bv{K}}{(2\pi)^3}
 \frac{\sin{K}}{K} 
\int_0^{\infty} dY 
\left ( 
\frac{\pi K_Z}{2K_\perp} + \frac{|K_x|}{K} M(\bv{K}) 
\right )^2 Y
e^{-  2\nu^*_2(0)K_x^2 Y^{3}/3}.
\end{eqnarray}
Hence, we find that the asymptotic form of $\bar{\Delta}_4(r)$ is given
by eq. (\ref{Delta4:asym}).

\end{document}